%% file: 0_AxOSyn_arxiv.tex
\documentclass{article}

\usepackage{arxiv}
\usepackage{graphicx}
\usepackage{textcomp}
\usepackage{amsmath,amsfonts,amssymb}
\usepackage{algorithm,algorithmic}
\usepackage{graphicx}
\usepackage[pscoord]{eso-pic}
\usepackage{lipsum}
\usepackage{booktabs} 
\usepackage{enumitem}
\usepackage{colortbl} 
\usepackage{multirow}
\usepackage{epstopdf}
\usepackage{diagbox}
\usepackage{svg}
\usepackage{xcolor}
\usepackage{verbatim} 
\usepackage{multicol}
\usepackage{hyphenat}
\usepackage{lipsum}
\usepackage{enumitem}
\usepackage{etoolbox}
\usepackage{mathtools}
\usepackage{euscript}
\usepackage{mathrsfs}
\usepackage{bm}
\usepackage[english]{babel}
\usepackage{blindtext}
\usepackage{textcase}
\usepackage{color}
\usepackage{tikz}
\usepackage{pifont}
\usepackage{makecell}
\usepackage[T1]{fontenc}
\usepackage{hyperref}
\usepackage[acronym]{glossaries}
\usepackage{subcaption}
\usepackage[tablename=TABLE, tableposition=top]{caption}
\usepackage{pifont}
\usepackage{makecell}
\usepackage[T1]{fontenc}
\usepackage{hyperref}
\usepackage[acronym]{glossaries}
\usepackage[nodayofweek]{datetime}
\usepackage[compact]{titlesec}

\newcommand{\cmark}{\ding{51}}%
\newcommand{\xmark}{\ding{55}}%

\DeclareMathAlphabet{\mathpzc}{OT1}{pzc}{m}{it}

\definecolor{add}{rgb}{1, 0, 0}
\definecolor{rev}{rgb}{0, 0, 1}
\definecolor{reg}{rgb}{0, 0, 0}

\addto\extrasenglish{%

}
\makeglossaries
\input{relatedAcronyms}
\input{01_docVars}

\title{\doctitle}

\date{} 					

\author{ \href{https://orcid.org/0000-0002-2243-5350}{\includegraphics[scale=0.06]{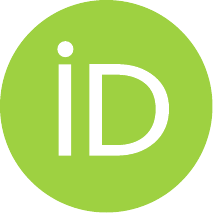}\hspace{1mm}Siva Satyendra Sahoo}\\
	\texttt{Siva.Satyendra.Sahooo@imec.be} \\
	\And
	\href{https://orcid.org/0000-0002-9774-9522}{\includegraphics[scale=0.06]{orcid.pdf}\hspace{1mm}Salim Ullah} \\
	\texttt{Salim.Ullah@rub.de} \\
	\And
	\href{https://orcid.org/0000-0001-7125-1737}{\includegraphics[scale=0.06]{orcid.pdf}\hspace{1mm}Akash Kumar} \\
	\texttt{Akash.Kumar@rub.de} \\
}

\renewcommand{\headeright}{Under Review: \textit{ACM TRETS}}
\renewcommand{\undertitle}{Under Review: \textit{ACM TRETS}}
\renewcommand{\shorttitle}{AxOSyn}

\begin{document}
	\glsresetall


\maketitle
\begin{abstract}
  \input{01_abstract}
\end{abstract}




\glsresetall
\input{02_manuscript_top}

\clearpage
\bibliographystyle{ACM-Reference-Format}
\bibliography{references.bib}

\end{document}

%% file: relatedAcronyms.tex
\newacronym{axc}{AxC}{\text{A}pproximate \text{C}omputing}
\newacronym{axo}{AxO}{\text{A}pproximate Arithmetic \text{O}perator}
\newacronym{ai}{AI}{\text{A}rtificial \text{I}ntelligence}
\newacronym{cpd}{CPD}{\text{c}ritical \text{p}ath \text{d}elay}
\newacronym{pdp}{PDP}{\text{P}ower-\text{D}elay \text{P}roduct}
\newacronym{ilp}{ILP}{\text{I}nstruction \text{L}evel \text{P}arallelism}
\newacronym{vlsi}{VLSI}{\text{V}ery \text{L}arge \text{S}cale \text{I}ntegration}
\newacronym{mttf}{MTTF}{\text{M}ean \text{T}ime \text{T}o \text{F}ailure}
\newacronym{mttc}{MTTC}{\text{M}ean \text{T}ime \text{T}o \text{C}rash}
\newacronym{noc}{NoC}{\text{N}etwork-on-\text{C}hip}
\newacronym{pr}{PR}{\text{P}olynomial \text{R}egression}
\newacronym{cots}{COTS}{commercial-off-the-shelf}
\newacronym{nre}{NRE}{\text{N}on \text{R}ecurring \text{E}ngineering}
\newacronym{rtl}{RTL}{\text{R}egister \text{T}ransfer \text{L}evel}
\newacronym{vcu}{VCU}{\text{V}ideo \text{C}odec \text{U}nit}
\newacronym{apu}{APU}{\text{A}pplication \text{P}rocessing \text{U}nit}
\newacronym{gpu}{GPU}{\text{G}raphics \text{P}rocessing \text{U}nit}
\newacronym{rpu}{RPU}{\text{R}eal-time \text{P}rocessing \text{U}nit}
\newacronym{gps}{GPS}{\text{G}lobal \text{P}ositioning \text{S}ystem}
\newacronym{ml}{ML}{\text{M}achine \text{L}earning}
\newacronym{iot}{IoTs}{\text{I}nternet of \text{T}hings}
\newacronym{ic}{ICs}{\text{i}ntegrated \text{c}ircuits}
\newacronym{esl}{ESL} {\text{E}lectronic \text{S}ystem \text{L}evel}
\newacronym{eda}{EDA} {\text{E}lectronic \text{D}esign \text{A}utomation}
\newacronym{qos}{QoS} {\text{Q}uality of \text{S}ervice}
\newacronym{soc}{SoC} {\text{S}ystem-\text{o}n-\text{C}hip}
\newacronym{fpga}{FPGA} {\text{F}ield \text{P}rogrammable \text{G}ate \text{A}rray}
\newacronym{dpr}{DPR} {\text{D}ynamic \text{P}artial \text{R}econfiguration}
\newacronym{prr}{PRR} {\text{P}artially \text{R}econfigurable \text{R}egion}
\newacronym{prm}{PRM} {\text{P}artially \text{R}econfigurable \text{M}odule}
\newacronym{pe}{PE} {\text{P}rocessing \text{E}lement}
\newacronym{dse}{DSE} {\text{D}esign \text{S}pace \text{E}xploration}
\newacronym{ga}{GA} {\text{G}enetic \text{A}lgorithms}
\newacronym{bti}{BTI} {\text{B}ias \text{T}emperature \text{I}nstability}
\newacronym{nbti}{NBTI} {\text{N}egative \text{B}ias \text{T}emperature \text{I}nstability}
\newacronym{pbti}{PBTI} {\text{P}ositive \text{B}ias \text{T}emperature \text{I}nstability}
\newacronym{em}{EM} {\text{E}lectro\text{m}igration}
\newacronym{gob}{GOB} {\text{G}ate \text{O}xide \text{B}reakdown}
\newacronym{hci}{HCI} {\text{H}ot \text{C}arrier \text{I}njection}
\newacronym{tddb}{TDDB}{\text{T}ime \text{D}ependent \text{D}ielectric \text{B}reakdown}
\newacronym{seu}{SEU} {\text{S}ingle \text{E}vent \text{U}pset}
\newacronym{ser}{SER} {\text{S}oft \text{E}rror \text{R}ate}
\newacronym{gdb}{GDB} {\text{G}ate \text{D}ielectric \text{B}reakdown}
\newacronym{tmr}{TMR} {\text{T}riple \text{M}odular \text{R}edundancy}
\newacronym{dmr}{DMR} {\text{D}ual \text{M}odular \text{R}edundancy}
\newacronym{ecc}{ECC}{\text{E}rror \text{C}hecking and \text{C}orrecting}
\newacronym{sram}{SRAM}{\text{S}tatic \text{R}andom \text{A}ccess \text{M}emory}
\newacronym{dram}{DRAM}{\text{D}ynamic \text{R}andom \text{A}ccess \text{M}emory}
\newacronym{llc}{LLC}{\text{L}ast \text{L}evel \text{C}ache}
\newacronym{l1}{L1}{\text{L}evel \text{1}}
\newacronym{dimm}{DIMM}{\text{D}ual \text{i}n-line-\text{M}emory \text{M}odule}
\newacronym{snc}{SNC}{\text{S}ingle-\text{N}ibble-error-\text{C}orrecting}
\newacronym{dnd}{DND}{\text{D}ouble-\text{N}ibble-error-\text{D}etecting}
\newacronym{sec}{SEC}{\text{S}ingle-bit-\text{E}rror-\text{C}orrecting}
\newacronym{ded}{DED}{\text{D}ouble-bit-\text{E}rror-\text{D}etecting}
\newacronym{dec}{DEC}{\text{D}ouble-bit-\text{E}rror-\text{C}orrecting}
\newacronym{ted}{TED}{\text{T}riple-bit-\text{E}rror-\text{D}etecting}
\newacronym{ivi}{IVI}{\text{I}nstruction \text{V}ulnerability \text{I}ndex}
\newacronym{fvi}{FVI}{\text{F}unction \text{V}ulnerability \text{I}ndex}
\newacronym{sed}{SED}{\text{S}obel \text{E}dge \text{D}etection}
\newacronym{cnn}{CNN}{\text{C}onvolutional \text{N}eural \text{N}etworks}
\newacronym{dnn}{DNN}{\text{D}eep \text{N}eural \text{N}etworks}
\newacronym{os}{OS}{\text{O}perating \text{S}ystem}
\newacronym{avf}{AVF}{\text{A}rchitectural \text{V}ulnerability \text{F}actor}
\newacronym{milp}{MILP}{\text{M}ixed \text{I}nteger \text{L}inear \text{P}rogramming}
\newacronym{miqcp}{MIQCP}{\text{M}ixed \text{I}nteger \text{Q}uadratically \text{C}onstrained \text{P}rogramming}
\newacronym{sofr}{SOFR}{\text{S}um-\text{o}f-\text{F}ailure \text{R}ate}
\newacronym{clb}{CLB}{\text{C}onfigurable \text{L}ogic \text{B}locks}
\newacronym{bram}{BRAM}{\text{B}lock \text{RAM}}
\newacronym{dsps}{DSPs}{\text{D}igital \text{S}ignal \text{P}rocessing blocks}
\newacronym{mcts}{MCTS}{\text{M}onte \text{C}arlo \text{T}ree \text{S}earch}
\newacronym{ttp}{TTP} {\text{T}ree \text{T}raversal \text{P}roblem}
\newacronym{fir}{FIR} {\text{F}inite \text{I}mpluse \text{R}esponse}
\newacronym{mtbf}{MTBF}{Mean Time between Failures}
\newacronym{ura}{\textit{uRA}}{User-modulated Run-time Adaptation}
\newacronym{aura}{\textit{AuRA}}{Agent-based User-modulated Run-time Adaptation}
\newacronym{moea}{MOEA}{\text{M}ulti-\text{O}bjective \text{E}volutionary \text{A}lgorithms}
\newacronym{dvfs}{DVFS}{\text{D}ynamic \text{V}oltage and \text{F}requncy \text{S}caling}
 \newacronym{icap}{ICAP}{\text{I}nternal \text{C}onfiguration \text{A}ccess \text{P}ort}
\newacronym{rl}{RL}{\text{R}einforcement \text{L}earning}
\newacronym{pvt}{PVT}{\text{P}rocess, \text{V}oltage, and \text{T}emperature}
\newacronym{nhpp}{NHPP}{\text{N}on-\text{H}omogeneous \text{P}oisson \text{P}rocess}
\newacronym{fit}{FIT}{\text{F}ailures \text{I}n \text{T}ime}
\newacronym{mosfet}{MOSFET}{\text{M}etal \text{O}xide \text{S}emiconductor \text{F}ield \text{E}ffect \text{T}ransistor }
\newacronym{nmos}{NMOS}{\text{N}egative channel \text{M}etal \text{O}xide \text{S}emiconductor}
\newacronym{pmos}{PMOS}{\text{P}ositive channel \text{M}etal \text{O}xide \text{S}emiconductor}
\newacronym{osi}{OSI}{\text{O}pen \text{S}ystems \text{I}nterconnection}
\newacronym{bist}{BIST}{\text{B}uilt-\text{i}n \text{S}elf \text{T}est}
\newacronym{nlp}{NLP}{\text{N}atural \text{L}anguage \text{P}rocessing}
\newacronym{dof}{DoF}{\text{D}egree \text{o}f \text{F}reedom}
\newacronym{lut}{LUT}{Look-Up Table}
\newacronym{ppa}{PPA}{Power-Performance-Area}
\newacronym{behav}{BEHAV}{behavioral accuracy}
\newacronym{mse}{MSE}{Mean Squared Error}
\newacronym{mlp}{MLP}{Multi-Layer Perceptron}
\newacronym{l2}{L2}{Squared Error}
\newacronym{ppf}{PPF}{Pseudo Pareto-front}
\newacronym{vpf}{VPF}{Validated Pareto-front}
\newacronym{rmse}{RMSE}{Root Mean Squared Error}
\newacronym{mac}{MAC}{Multiply-Accumulate}
\newacronym{asic}{ASIC}{Application-specific Integrated Circuit}
\newacronym{gan}{GAN}{Generative Adversarial Network}
\newacronym{ann}{ANN}{Artificial Neural Network}
\newacronym{aiml}{AI/ML}{Artifical Intelligence and Machine Learning}
\newacronym{lpf}{LPF}{Low-pass Filter}
\newacronym{ecg}{ECG}{Electrocardiogram}
\newacronym{shap}{SHAP}{SHapley Additive exPlanations}
\newacronym{bfs}{BFS}{Breadth First Search}
\newacronym{dfs}{DFS}{Depth First Search}
\newacronym{csp}{CSP}{Constraint Satisfaction Problem}
\newacronym{cgp}{CGP}{Cartesian Genetic Programming}
\newacronym{cc}{CC}{Carry-propagation-Chain}
\newacronym{conss}{ConSS}{Configuration Supersampling}
\newacronym{dtr}{DTR}{Decision Tree Regression}
\newacronym{rfr}{RFR}{Random Forest Regression}
\newacronym{gbm}{GBM}{Gradient Boosting Machine}
\newacronym{nn}{NN}{Neural Network}
\newacronym{axat}{AxAT}{Approximation-aware Training}
\newacronym{mbo}{MBO}{Multi-objective Bayesian Optimization}
\newacronym{gnn}{GNN}{Graph Neural Network}

%% file: 01_docvars.tex
\newcommand{\doctitle}{
    \textit{AxOSyn}: An Open-source Framework for Synthesizing Novel Approximate Arithmetic Operators
    }

\newcommand{\titleName}{\textit{AxOSyn}}
\newcommand\mylistkeywords{
Approximate Computing,
Design Space Exploration,
Open-source Tools,
Electronic Design Automation,
AI/ML for EDA,
}

\newcommand{\siva}[1]{\textcolor{black}{#1}}

%% file: 01_abstract.tex
\siva{
Edge AI deployments are becoming increasingly complex, necessitating energy-efficient solutions for resource-constrained embedded systems.
Approximate computing, which allows for controlled inaccuracies in computations, is emerging as a promising approach for improving power and energy efficiency. 
Among the key techniques in approximate computing are approximate arithmetic operators (AxOs), which enable application-specific optimizations beyond traditional computer arithmetic hardware reduction-based methods, such as quantization and precision scaling.
Existing design space exploration (DSE) frameworks for approximate computing limit themselves to selection-based approaches or custom synthesis at fixed abstraction levels, which restricts the flexibility required for finding application-specific optimal solutions. 
Further, the tools available for the DSE of  AxOs are quite limited in terms of exploring different approximation models and extending the analysis to different granularities.
To this end, we propose \textit{AxOSyn}, an open-source framework for the DSE of AxOs that supports both selection and synthesis approaches at various abstraction levels. AxOSyn allows researchers to integrate custom methods for evaluating approximations and facilitates DSE at both the operator-level and application-specific. Our framework provides an effective methodology for achieving energy-efficient, approximate operators.
}

%% file: 02_manuscript_top.tex
\input{10_intro}        
\clearpage
\input{20_bckrel}       
\clearpage
\input{30_tool_framework}     
\input{40_tool_features}  
\newpage
\input{50_expres}       
\input{60_conc}         

%% file: 10_intro.tex
\section{Introduction}
\label{sec:intro}
\siva{
\gls{aiml} models are increasingly being deployed in edge devices across a variety of applications, often with growing complexity~\cite{10.1145/3724420}.
Achieving scalable performance on these resource-constrained devices becomes more challenging as the models become more demanding in terms of both computation and memory requirements~\cite{9985008}.
To address these challenges, energy-efficient computing has become crucial for enabling practical AI inference at the edge, where power and computational resources are limited. 
Application-specific optimizations, such as pruning, knowledge distillation, and quantization, are often employed to achieve this energy efficiency. 
These techniques help to tailor AI models to the capabilities of the target hardware, reducing power and memory requirements while maintaining acceptable performance.
\glspl{fpga} have emerged as a promising platform for deploying edge AI due to their low power consumption and inherent flexibility. 
Their ability to accelerate the diverse and dynamic nature of models being used today makes them ideal for energy-efficient edge AI deployments. 
In addition to the aforementioned optimization techniques, \gls{axc} has also emerged as a potentially effective approach to enhancing energy efficiency. 
By allowing for controlled errors or approximations in arithmetic operations, \gls{axc} can provide significant gains in power efficiency, making it a valuable tool for AI computing in resource-constrained environments.
\gls{axc} has been witnessing increasing interest in the field of AI computing, as it allows the designers to leverage the implicit error tolerance of such algorithms.
}

\siva{
\gls{axc} comprises various methods deployed at different layers of the computing stack -- application and system software and hardware~\cite{10.1145/2893356}.
In terms of hardware-based approximations, one of the more promising approaches involves using \glspl{axo}. 
Since most of the processing in AI inference involves computer arithmetic, reducing the implementation cost for each such processing can result in a large overall energy reduction.
Approximate arithmetic goes beyond traditional quantization or precision scaling, offering greater scope for application-specific optimizations. 
Designing application-specific \glspl{axo} can be approached in two distinct ways: \textit{selection} and \textit{synthesis}. 
\textit{Selection}, the first approach, involves choosing the most suitable \glspl{axo} from a pre-existing set of operators that have been explored and characterized at the operator level. 
This allows designers to match the operators to the specific requirements of an application and keeps the design space relatively manageable.
}

\begin{figure}[t] 
	\centering
	\subfloat[]{
		\centering
		\scalebox{1}{\includegraphics[width=0.55 \columnwidth]{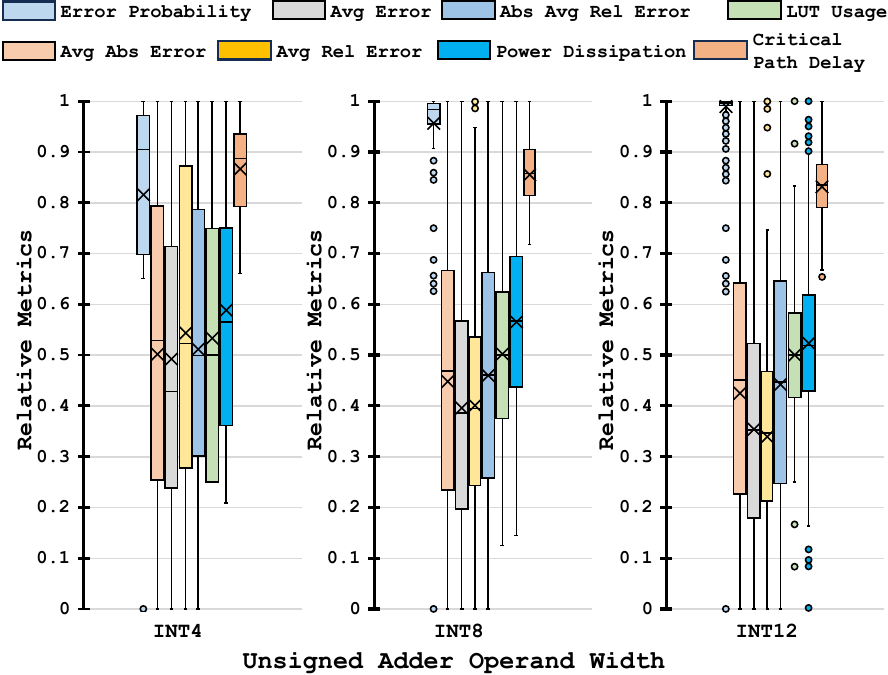}}
	}
	\subfloat[]{
        \centering
		\scalebox{1}{\includegraphics[width=0.45 \columnwidth]{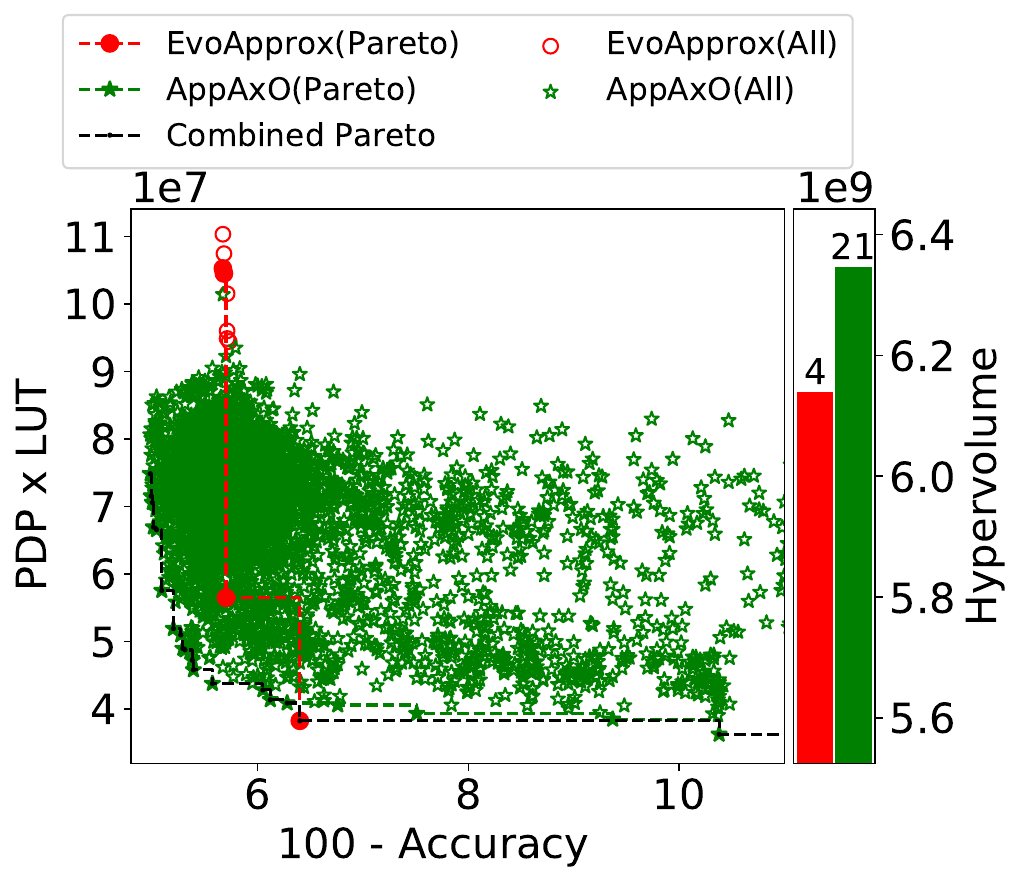}}
	}
	\caption{
    Approximate arithmetic operators (a)~Distribution of error and PPA metrics of 6- and 8-bit approximate signed adders (b)~Improved error-PPA trade-offs with application-specific signed multiplier AxO search~\cite{ullah2022appaxo}
    }
	\label{fig:mot_appaxo} 
\end{figure}

\siva{
\textit{Synthesis}, on the other hand, involves creating novel, application-specific \glspl{axo} rather than relying solely on an existing set of \glspl{axo}. 
Since it usually involves some form of automated synthesis of candidate \glspl{axo} designs, it enables generating a library of operator implementations. 
For instance, \autoref{fig:mot_appaxo}(a) illustrates the range of metrics for unsigned approximate adders of various bit widths as box plots, showcasing a large number of designs with varying trade-offs obtained through the synthesis approach.
The designs were generated using the proposed \titleName~framework while adopting the operator approximation model presented in~\cite{ullah2022appaxo}. 
The plots show the relative range of both \gls{behav} and \gls{ppa} metrics for $15$, $255$, and $4095$ approximate implementations for the unsigned INT4, INT8, and INT12 adders, respectively.
The synthesis approach opens up possibilities for more tailored optimizations that directly meet the needs of the target application. 
However, both approaches result in large design spaces that need to be explored to extract the maximum benefits through \gls{axc}.
}

\siva{
Various related works have proposed novel methods to effectively traverse these large design spaces. 
For instance, \autoref{fig:mot_appaxo}(b) shows the result of the \gls{dse} for finding the 8-bit signed integer multiplier \glspl{axo} in the \gls{lpf} of the pre-processing stage of ECG classification.
\autoref{fig:mot_appaxo}(b) shows the design points obtained through the synthesis-based methodology of AppAxO~\cite{ullah2022appaxo} and the selection-based method using the multipliers in EvoApprox~\cite{evoapprox16}.
As seen from the figure, the synthesis approach offers a much larger and more diverse set of potential designs, highlighting its promise for application-specific optimizations. 
This is evident from the comparison of the Pareto front design points and the resulting hypervolume shown in the figure.
However, using a synthesis-based approach also means that a large number of designs need to be evaluated to identify the \gls{axo} implementations that offer the optimal \gls{behav}-\gls{ppa} trade-offs. 
Various works have tried to get around the large design space for both synthesis and selection-based approaches by using evolutionary optimization methods~\cite{ullah2022appaxo,evoapprox16,9072581,10.1145/3648694,10504662} and \gls{aiml}-based search optimizations~\cite{10.1145/3566097.3567891,ullah2022appaxo,10504662,10.1145/3609096}.
}

\siva{
However, existing frameworks for \gls{dse} often either focus on selection-based approaches or rely on custom synthesis at specific abstraction levels, limiting the ability to explore novel approximation models and varying levels of abstraction. 
Furthermore, very few frameworks provide the flexibility to explore new synthesis methods for \glspl{axo} effectively. 
To this end, we present \titleName, an open-source framework for efficient \gls{dse} of novel approximate operators, enabling users to explore custom synthesis approaches and fully leverage the benefits of \gls{axc}.
The related contributions are listed below.\\
}
\siva{
\textbf{Contributions:}
\begin{enumerate}
    \item The proposed framework includes a separation of functionality and hardware implementation that allows the exploration of \glspl{axo} at varying abstraction levels. Further, \titleName~allows the integration of custom methods of evaluating the effect of approximations and approximation models.
    \item The proposed framework integrates \gls{dse} methods for both selection and synthesis approaches. Further, \titleName~allows the \gls{dse} at both the operator and application levels. In addition to integrating some of the more popular methods for \gls{dse}, the separation of design and \gls{dse} components allows easy integration of custom search algorithms.
    \item The proposed framework is open source and available at \href{https://blinded\_for\_review}{https://blinded\_for\_review}.
\end{enumerate}
}

\siva{
The rest of the article is organized as follows.
Section~\ref{sec:bckrel} presents a brief background on \gls{axc} and the implementation of \glspl{axo}. The section also includes a brief survey of related works.
The proposed \titleName~framework is presented in Section~\ref{sec:tool_arch}. It includes a detailed description of the various components of the framework and how they are integrated into \titleName.
Section~\ref{sec:tool_features} presents the different features of \titleName. The core features, the user interface, and the extensibility of \titleName~ are discussed in this section.
Section~\ref{sec:expres} presents some results obtained using the proposed framework and the related discussion. The results are used to demonstrate the usefulness of the various features of \titleName.
The article is concluded in Section~\ref{sec:conc} with a summary and a brief discussion of related future research, including the proposed extensions to \titleName.
}

%% file: 20_bckrel.tex
\section{Background and Related Works}
\label{sec:bckrel}

\input{21_axc}

\newpage

\input{22_axo_dse}

\input{23_table_summary}

\subsection{Summary}

\siva{
Based on the above discussion, the implementation of \glspl{axo} that provide the optimal \gls{behav}-\gls{ppa} trade-offs involves various deign and engineering aspects.
\autoref{table:compare_rel} shows a categorization of these aspects, along with how some of the related works have contributed to them.
The selection/synthesis approach to implementing constrained \gls{axo}-based designs presents varying degrees of design space size and scope for targeted optimizations.
Similarly, application-specific \gls{dse} forms a major research goal for \gls{axo} design since it can allow for fine-grained optimizations tailored for each application.
However, increasing the scope of optimizations also leads to large design spaces, where modern advancements in \gls{aiml} can be leveraged for \gls{dse}.
One of the more frequent uses of \gls{aiml} in \gls{dse} involves the regression models-based prediction of \gls{behav}/\gls{ppa} metrics rather than costly physical characterization. 
Finally, extensibility forms an important aspect of \gls{dse} frameworks as it allows researchers to experiment with novel designs, models, and search methods.
}

\siva{
While various research works have enabled the exploration of \glspl{axo} from one or more aspects, most works are limited in terms of their extensibility or do not provide an open-source framework to integrate custom methods.
To this end, we present \titleName, an open source framework to empower researchers for exploring novel approaches to \gls{axc} in general and \glspl{axo} in particular. 
To the best of our knowledge, no prior work integrates all the design aspects shown in~\autoref{table:compare_rel} in a joint framework to enable an efficient search for \glspl{axo}.
}

%% file: 21_axc.tex
\begin{figure}[t] 
    \centering
       \scalebox{1}{\includegraphics[width=1 \columnwidth]{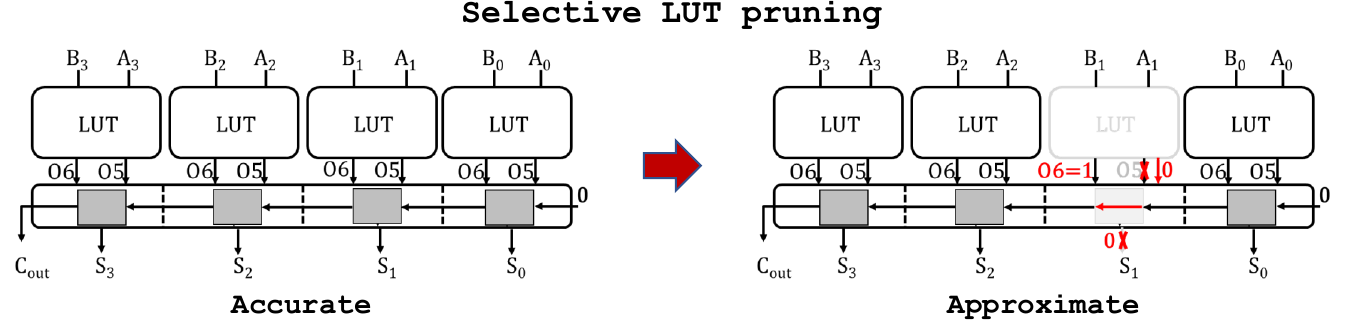}}
  \caption{Approximate arithmetic operator design in FPGAs using selective LUT-pruning~\cite{ullah2022appaxo}}
  \label{fig:bcklit_axo} 
\end{figure}
\subsection{Approximate Computing}
\siva{
\acrfull{axc} has emerged as a promising solution for managing the increasing computational and memory demands of emerging applications, particularly those that demonstrate resilience to minor inaccuracies. 
These error-tolerant applications, such as AI computing, can generate multiple acceptable outputs rather than a single precise result, which \gls{axc} leverages to offer trade-offs between accuracy and performance. 
Approximate techniques span multiple layers of the computing stack, with significant attention given to the architecture and circuit layers for resource-constrained, compute-intensive applications. 
At the architectural level, reduced-precision arithmetic and hybrid quantization techniques have proven effective for reducing computational and memory overheads.
Similarly, circuit-level approximations often involve using simplified computational units such as approximate adders and multipliers~\cite{10.1145/2893356,chippa2014scalable,yin2017minimizing,miguel2015doppelganger,venkataramani2015computing,Ullah2023,wang2019bfloat16,9492075}. 
These approaches are particularly valuable for implementing \gls{aiml} inference on embedded systems where power and energy constraints are critical.
}


\siva{
Among the various approaches to the deliberate introduction of inaccuracies for implementing \glspl{axo}, most methods can be categorized into one of two types: \textit{truncation} or \textit{composition}. 
Truncation involves selectively removing parts of the computational structures to introduce approximations, such as truncating the carry propagation chain in adders or selectively omitting critical \glspl{lut} in multipliers~\cite{6691096,8342140,5771062,9072581,7827657}. 
For instance, \autoref{fig:bcklit_axo} shows the generation of an approximate 4-bit unsigned adder by pruning the second \gls{lut} that is involved in generating the second \textit{sum} bit from the LSB.
With the truncation approach, the determination of the right subset of logic that can be pruned poses a significant research problem and has led to multiple novel approaches being proposed in literature. 
For instance, the authors of~\cite{9072581} used the accurate operator implementation to rank the top power consuming \glspl{lut} and critical paths. 
The ranking was then used to remove \glspl{lut}, along with some modification to the logic of the remaining \glspl{lut} to recover some level of error.
The composition approach, on the other hand,  focuses on building larger operators from smaller approximate units, thereby creating more complex operators with specific trade-offs~\cite{10.1145/3195970.3196115}. 
Multiple recent works have explored \gls{asic}- and \gls{fpga}-based implementations of approximate adders and multipliers using diverse techniques such as \gls{cgp} to generate optimized designs~\cite{hashemi2015drum,9233379,5771062}. 
These designs result in varying levels of trade-offs between power, area, and accuracy, underscoring the need for systematic exploration of design options to effectively exploit \glspl{axo} in specific application domains. 
In this work, we focus on the truncation approach to synthesize novel approximate arithmetic operators, aiming to explore the extensive design space of \glspl{axo} for efficient, application-specific optimizations.
However, it must be noted that the modular approach of \titleName~can be used to include the analysis and optimization with the composition approach.
}

%% file: 22_axo_dse.tex
\begin{figure}[t] 
    \centering
       \scalebox{1}{\includegraphics[width=0.8 \columnwidth]{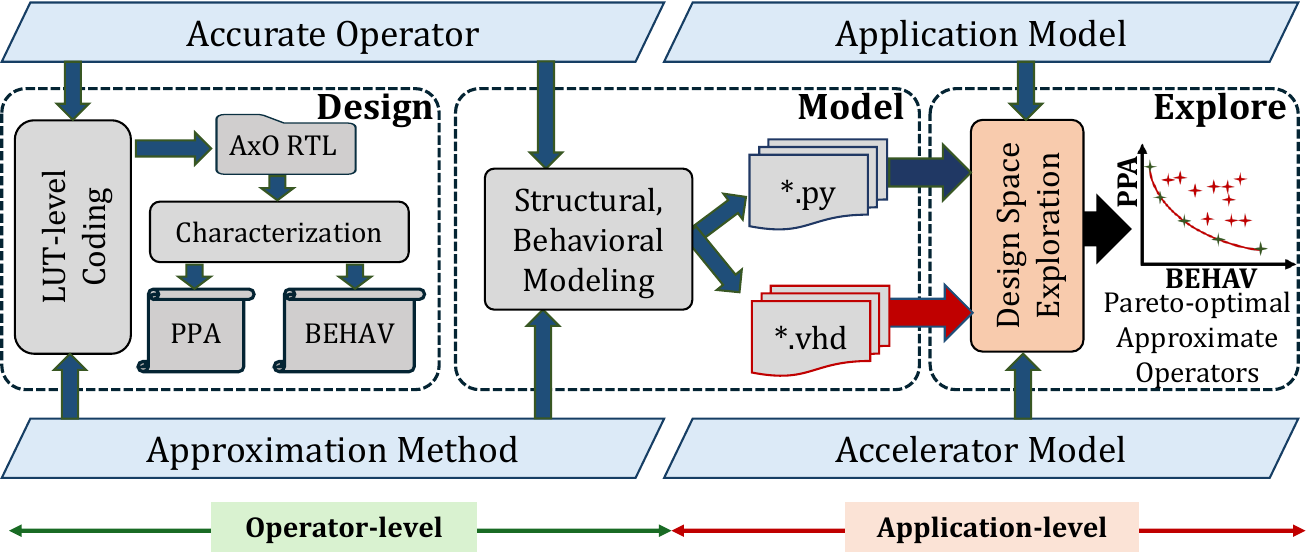}}
  \caption{Generic methodology for designing approximate operators~\cite{10740734}}
  \label{fig:meth_axo} 
\end{figure}
\subsection{DSE for Approximate Operators}
\siva{
One of the primary approaches to extracting the maximum benefits from using \glspl{axo} involves enabling platform-specific optimizations during the design and implementation.
While most of the existing works on \glspl{axo} focus on \gls{asic} implementations~\cite{evoapprox16}, some have explored \gls{fpga}-specific designs utilizing \glspl{lut} and \glspl{cc} structures to propose various approximate architectures~\cite{9344673,10.1145/3195970.3196115,9072581}. 
On the other hand, some related works adopt a more \textit{trasfer}-based approach by implementing \gls{asic}-optimized \glspl{axo} on an \gls{fpga}~\cite{mrazek2019autoax,9218533}.
Irrespective of minor differences in the approach, the design of novel \glspl{axo} usually involves some common aspects.
\autoref{fig:meth_axo} shows a generic methodology that shows how these aspects are involved in the search for novel \glspl{axo}~\cite{10740734}.
The \textit{design} aspect involves generating the platform-optimized implementation of an accurate operator.
The \textit{modeling} aspect involves using an approximation model, such as selective \gls{lut} pruning, with the accurate operator implementation to generate candidate \glspl{axo} designs.
The \textit{exploration} aspect usually involves using search methods that include the application behavior to find the Pareto front \glspl{axo} for an application.
}
%

\siva{
Various works have adopted a more \textit{manual} \gls{dse} approach to designing novel \glspl{axo}, resulting in more application-agnostic designs~\cite{9072581}. 
For example, \cite{9344673} and \cite{8465781} presented $4\times4$ approximate multipliers, which were extended to create larger multipliers, whereas \cite{10.1145/3195970.3196115} used different approximate $4\times4$ multipliers to form a higher-order multiplier library.
However, the best use of a methodology such as the one shown in~\autoref{fig:meth_axo} would require automated methods for one/more of the design, modeling, and exploration stages.
Consequently, some recent studies have started using automated \gls{dse} techniques to generate libraries with numerous approximate versions of operators, balancing accuracy and performance trade-offs~\cite{9218533,mrazek2019autoax,ullah2022appaxo}. 
These methods often involve creating models of arithmetic circuits to generate and evaluate novel configurations systematically. 
However, some of these approaches may limit the \gls{dse} to selecting designs from pre-existing \gls{asic}-optimized logic libraries, leading to constrained exploration~\cite{9233379,9218533,mrazek2019autoax}. 
Automated frameworks, such as \cite{10.1145/2966986.2967021}, have used \gls{cgp} for designing \gls{asic}-optimized approximate multipliers but lacked consideration for key metrics like dynamic power consumption. 
Similarly, frameworks like \cite{ullah2022appaxo,10.1145/3583781.3590222} that allow pruning-based automatic design generation and synthesis limit the exploration to a single approximation model.
}

\siva{
In any of the automated \gls{dse} methods, the large design space and the characterization time for candidate designs form bottlenecks to finding optimal \glspl{axo}.
To tackle this challenge, researchers are increasingly looking to intelligent \gls{dse} methods that leverage \gls{ai}/\gls{ml} techniques to guide the search toward optimal solutions. 
This ranges from using \gls{ml}-based pseudo fitness functions~\cite{ullah2022appaxo,9218533,10.1145/3648694,10504662} to using more complex generative and explainable \gls{ai} for aiding the design search algorithms~\cite{10504662,10.1145/3566097.3567891,10.1145/3609096}.
For instance, the authors of \cite{ullah2022appaxo} proposed a \gls{ga}-based synthesis with ML-driven fitness evaluation, while \cite{10.1145/3566097.3567891} utilized \gls{gan} to discover operator configurations that meet specific accuracy and performance requirements.
Similarly, in~\cite{10.1145/3609096}, the authors have used the SHAP values from the regression models to predict the \gls{ppa} and \gls{behav} metrics to guide a vanilla \gls{mcts} algorithm searching for novel \glspl{axo}. 
Further, the authors of~\cite{10.1145/3648694} have used the regression models to formulate \gls{miqcp} problems that augment the initial population of \gls{ga}.
}

%% file: 23_table_summary.tex
\begin{table}[t]
\centering
\caption{Comparing related works}
\label{table:compare_rel}
\def\arraystretch{1.3}
\resizebox{1\textwidth}{!}{
\begin{tabular}{lcccccc}
\hline
\multicolumn{2}{c}{\textbf{Related Works}} & EvoApprox~\cite{evoapprox16} & ApproxFPGA~\cite{9218533} & AppAxO~\cite{ullah2022appaxo} & AGNApprox~\cite{trommer2022} & AxOSyn (proposed) \\ \hline
\textbf{Aspects} & \textit{\textbf{\begin{tabular}[c]{@{}c@{}}Main \\ Focus\end{tabular}}} & \textit{\begin{tabular}[c]{@{}c@{}}Novel ASIC-based\\ AxO generation\end{tabular}} & \textit{\begin{tabular}[c]{@{}c@{}}Porting ASIC\\ optimized designs \\ for FPGAs\end{tabular}} & \textit{\begin{tabular}[c]{@{}c@{}}ML-based DSE \\ with automated \\ LUT pruning\end{tabular}} & \textit{\begin{tabular}[c]{@{}c@{}}PyTorch-based \\ evaluation and \\ selection of AxOs\end{tabular}} & \textit{\begin{tabular}[c]{@{}c@{}}Holistic framework \\ with extensibility \\ across aspects\end{tabular}} \\ \hline
\multirow{2}{*}{\textbf{\begin{tabular}[c]{@{}l@{}}Operator\\ Generation\end{tabular}}} & \textit{\textbf{Selection}} & \xmark & \cmark & \xmark & \cmark & \cmark \\ \cline{2-2}
 & \textit{\textbf{Synthesis}} & \cmark & \xmark & \cmark & \xmark & \cmark \\ \hline
\multirow{2}{*}{\textbf{\begin{tabular}[c]{@{}l@{}}DSE \\ Approach\end{tabular}}} & \textit{\textbf{\begin{tabular}[c]{@{}c@{}}Application\\ specific\end{tabular}}} & \cmark & \cmark & \cmark & \cmark & \cmark \\ \cline{2-2}
 & \textit{\textbf{\begin{tabular}[c]{@{}c@{}}AI/ML\\ enabled\end{tabular}}} & \xmark & \cmark & \cmark & \xmark & \cmark \\ \hline
\multirow{2}{*}{\textbf{\begin{tabular}[c]{@{}l@{}}Metric\\ Estimation\end{tabular}}} & \textit{\textbf{Physical}} & \xmark & \cmark & \cmark & \xmark & \cmark \\ \cline{2-2}
 & \textit{\textbf{Low-cost}} & \cmark & \cmark & \cmark & \xmark & \cmark \\ \hline
\multirow{4}{*}{\textbf{\begin{tabular}[c]{@{}l@{}}Extensibility\\ of Primary\\ Features\end{tabular}}} & \textit{\textbf{AxO models}} & \xmark & \xmark & \xmark & \cmark & \cmark \\ \cline{2-2}
 & \textit{\textbf{DSE methods}} & \xmark & \xmark & \cmark & \xmark & \cmark \\ \cline{2-2}
 & \textit{\textbf{\begin{tabular}[c]{@{}c@{}}Estimation \\ Methods\end{tabular}}} & \xmark & \xmark & \xmark & \xmark & \cmark \\ \cline{2-2}
 & \textit{\textbf{Accelerators}} & \xmark & \cmark & \cmark & \xmark & \cmark \\ \hline
\end{tabular}
}
\end{table}

%% file: 30_tool_framework.tex
\section{\titleName~Framework}
\label{sec:tool_arch}
\begin{figure}[t] 
    \centering
       \scalebox{1}{\includegraphics[width=0.8 \columnwidth]{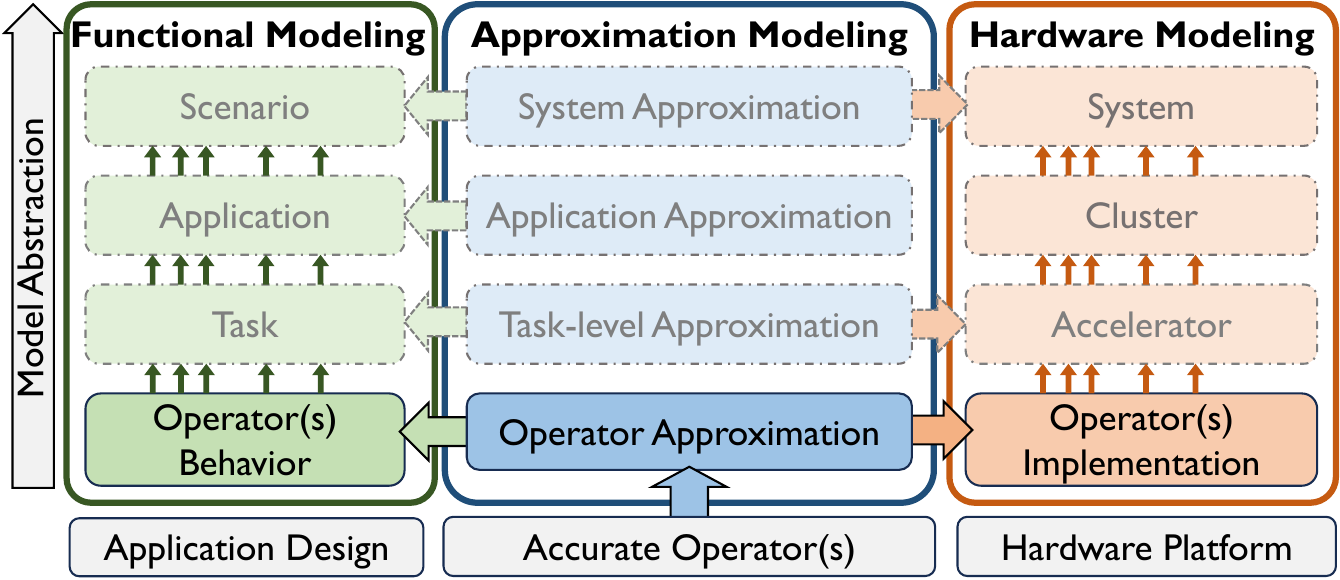}}
  \caption{Modeling of approximate components at different granularity}
  \label{fig:arch_model} 
\end{figure}

\siva{
\titleName~aims to combine the various aspects necessary for implementing and testing \glspl{axo} while providing sufficient abstraction to explore novel approaches. 
To this end, the proposed framework contains design, modeling, and \gls{dse} elements that can be used to implement novel methods for some components while using the existing methods for others. 
The different aspects of \titleName~along with the corresponding problem formulation for the \gls{axo} search are described next.
}

\input{32_ax_components}
\input{33_axo_models}
\begin{figure}[t] 
    \centering
       \scalebox{1}{\includegraphics[width=0.7 \columnwidth]{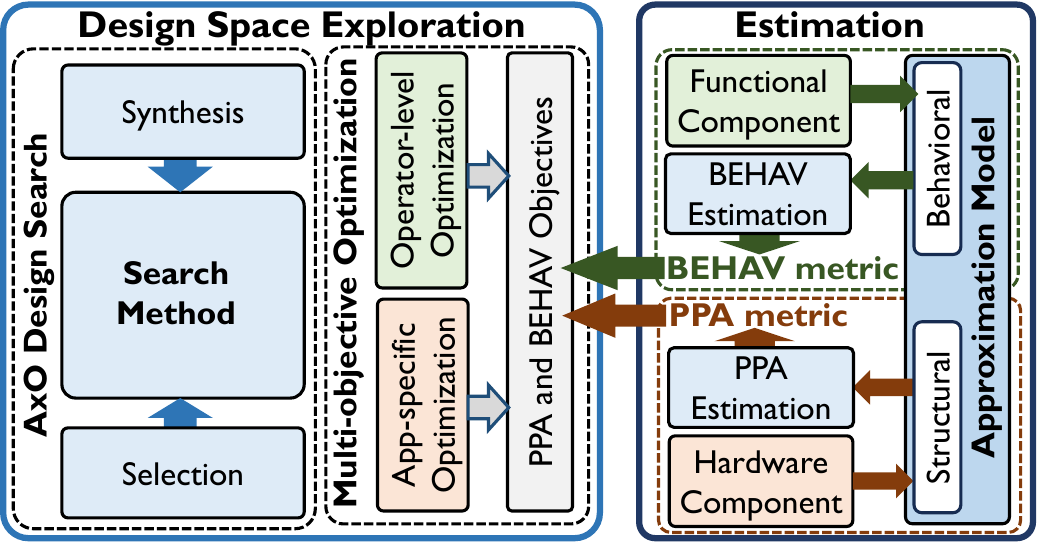}}
  \caption{The components of DSE for approximate arithmetic operators, including estimation methods of varying complexity. The various components interact with well-defined interfaces that allow novel approaches}
  \label{fig:arch_dse} 
\end{figure}
\newpage
\input{34_axo_dse}

%% file: 32_ax_components.tex
\subsection{Approximate Components}
\label{subsec:tool_frame_axcomp}
\siva{
The proposed framework uses the hierarchical primitives for the design components as shown in~\autoref{fig:arch_model}. 
The functional modeling assumes that a \textit{scenario} comprises different \textit{applications}, each in turn comprising different \textit{tasks}. 
The effect of using \glspl{axo} is included in the functional model of each task. This follows the assumption that the \gls{behav} effects of implementing different arithmetic operators are best captured in the resulting \gls{behav} of the task.
The hierarchical design and the resulting abstraction can allow the introduction of approximations beyond arithmetic operators with task/application/scenario level approximation models. 
In the current article, with the focus on approximate operators, we limit the investigation to the impact of \gls{axo} on operator-level behavior only.
The estimation of task/application-level behavior is achieved through the integration with other state-of-the-art tools.
}

\siva{
Similar to the \textit{Application Design}, the \textit{Hardware Platform} model assumes many-to-one mappings between \gls{axo} implementation and accelerators and multiple approximate accelerators in the system. 
It must be noted that we treat the accurate implementation of various components as part of the set of approximate implementations. 
Eq.~\eqref{eq:func_model} shows the mathematical formulation for the functional model for \textit{Application Design}.
The outputs of any task $t$ are a function of the behavior of the task $\mathcal{T}_t$ along with the functionality of the set of its constituent operators $\{\mathcal{O}_o\}$ given the set of inputs to the task $Inp_t$. 
For instance, the behavior of a neural network depends upon the type of its constituent layers (tasks) and the inputs to the application. The behavior of each layer, in turn, depends on the type of layer ($\mathcal{T}_t$), the set of arithmetic operators used in the computation $\{\mathcal{O}_o\}$, and the inputs to the layer. 
}
\begin{equation}
\label{eq:func_model}
\small{
    \begin{split}
        Out_{\mathcal{T}_t} = Out_{t,\{\mathcal{O}_o\}} &= \mathcal{T}_t (\{\mathcal{O}_o\}, Inp_t), ~~ for~any~task~t \\
        Out_{a,\{Out_{\mathcal{T}_t}\}_a} &= \mathcal{A}_a (\{Out_{t,\{\mathcal{O}_o\}}\}, Inp_a) ~~ for~an~app~a \\
        BEHAV_{\mathcal{T}_t,\{\mathcal{O}_o\}} &= Err\mathcal{T}_t(Out_{t,\{\mathcal{O}_o\}}, golden\_outputs)\\
        BEHAV_{\mathcal{A}_a,\{\mathcal{T}_t\}} &= Err\mathcal{A}_a(\{Out_{\mathcal{T}_t}\}_a, golden\_outputs)
    \end{split}
}
\end{equation}
\newpage
\begin{equation}
\label{eq:ppa_accel}
\small{
    \begin{split}
        PPA_{x} = \mathcal{{X}}_{x} (\{\mathcal{O}_x\}, \{uArch_x\},Inp_x), ~~ for~any~accelerator~x
    \end{split}
}
\end{equation}

\siva{
For the current article, we have limited the hardware modeling to the analysis of the operator only. 
However, the corresponding problem formulation for an accelerator using a set of \glspl{axo} is shown in Eq.~\eqref{eq:ppa_accel}.
We denote $PPA_{x}$ to denote any combination of the power dissipation, performance (\gls{cpd}), and area (\gls{lut} usage) metrics of an accelerator using the set of operators $\{\mathcal{O}_o\}$. 
In addition to $\{\mathcal{O}_x\}$, the accelerators' \gls{ppa} metrics also depend upon the architecture parameters, such as loop unrolling, pipelining, etc., denoted by $\{uArch_x\}$.
It can be noted that while the \gls{cpd} and \gls{lut} usage do not depend upon the inputs to the accelerator, the dynamic power does, and $Inp_x$ is hence included in the formulation.
}


%% file: 33_axo_models.tex
\subsection{Operator Approximation Models}
\label{subsec:tool_frame_axomodel}
\siva{
While the \titleName~framework can be extended to include the approximation at various layers, as shown in~\autoref{fig:arch_model}, we limit the current article to approximation modeling in arithmetic operators. 
As explained in~Section~\ref{sec:bckrel}, with the truncation-based approach, most \gls{axo} implementations are derived from some form of circuit pruning. 
Eq.~\eqref{eq:axo_model} captures this approach, where the accurate operator is represented by $\mathcal{O}_{ac}$ and the application of any arbitrary model $AxOM_m$ to $\mathcal{O}_{ac}$ using a model-specific approximate configuration $config_o$ results in $\mathcal{O}_{o}$. 
For models that start with an existing set of \gls{axo} designs, such as in~\cite{9218533,mrazek2019autoax}, the individual implementations and their behavior can be identified by an index from a table of values as shown in Eq.~\eqref{eq:axo_selection}. 
However, for models such as AppAxO~\cite{ullah2022appaxo,10.1145/3583781.3590222} that allow synthesis of novel \glspl{axo}, the implementation and behavior are captured by a binary string, as shown in Eq.~\eqref{eq:axo_synthesis}.
It must be noted that \titleName~can be used to integrate other models as long as specific interfaces for identification, functionality for a set of given inputs, and generating the RTL for physical characterization are provided.
}

\begin{equation}
\label{eq:axo_model}
\small{
    \begin{split}
        \mathcal{O}_o = AxOM_m(\mathcal{O}_{ac},config_o),& ~~ for~model~AxOM_m~and~config_o \\
        Out_{\mathcal{O}_{ac}} = \mathcal{F}_\mathcal{O}(\mathcal{O}_{ac},Inp); &~~~~ Out_{\mathcal{O}_{o}} = \mathcal{F}_\mathcal{O}(\mathcal{O}_{},Inp_{\mathcal{O}}) \\
        BEHAV_{\mathcal{O}_o} = & Err\mathcal{O}_e (Out_{\mathcal{O}_{ac}}, Out_{\mathcal{O}_{o}}) \\
        PPA_{\mathcal{O}_o} = & PPA\mathcal{O}_p (Out_{\mathcal{O}_{o}},Inp_{\mathcal{O}})
     \end{split}
}
\end{equation}
\begin{equation}
\label{eq:axo_selection}
\small{
    \begin{split}
        \mathcal{O}_{E} = \{ \mathcal{O}_l\}, where~l~=~index~of~list~of~AxOs
    \end{split}
}
\end{equation}
\begin{equation}
\label{eq:axo_synthesis}
\small{
    \begin{split}
        \mathcal{O}_{A} = \{ \mathcal{O}_i\}, where~\mathcal{O}_i=(l_0,l_1,..., l_i, ... l_L); l_i \in \{ 0,1\} 
    \end{split}
}
\end{equation}


%% file: 34_axo_dse.tex
\subsection{DSE for Approximate Operators}
\label{subsec:tool_frame_dse}
\siva{
The interaction of the various components of the \titleName~ framework for the \gls{dse} of \glspl{axo} are shown in~\autoref{fig:arch_dse}. 
\textit{AxO Design Search} involves finding the appropriate set of \glspl{axo} and includes using a specific search algorithm by implementing either synthesis or selection approaches. 
Abstracting the selection/synthesis approach from the search method allows the designer to experiment with different search algorithms.
Further, it also allows using the less complex selection approach even for synthesis-based operator models by using a subset of the designs explored via the synthesis approach.
The corresponding constrained optimization problems are shown in Eq.~\eqref{eq:dse_sel_syn}. 
${BEHAV}_{\mathcal{O}_o}$ and ${PPA}_{\mathcal{O}_o}$ denote any arbitrary behavioral error and \gls{ppa} metric respectively, or their combination thereof, when using the approximate operator $\mathcal{O}_o$.
Similarly, ${BEHAV}_{MAX}$ and ${PPA}_{MAX}$ denote the upper bound error and \gls{ppa} metrics respectively.
}

\begin{equation}
\label{eq:dse_sel_syn}
\small{
\begin{split}
    \underset{\mathcal{O}_o \in \mathcal{O}}{\text{minimize}} &({BEHAV}_{\mathcal{O}_o},{PPA}_{\mathcal{O}_o}) \\
    s.t. ~ {BEHAV}_{\mathcal{O}_o} &\leq BEHAV_{MAX} ~~ and \\
    {PPA}_{\mathcal{O}_o} &\leq PPA_{MAX} \\
    \mathcal{O}:~Set~of~all~possible~AxOs. &~Selection:~\mathcal{O}_E;~Synthesis:~\mathcal{O}_A 
\end{split}
}
\end{equation}

\siva{
The multi-objective optimization shown in Eq.~\eqref{eq:dse_sel_syn} denotes operator-level \gls{dse}. 
Similarly, the application-specific DSE problem is shown in Eq.~\eqref{eq:dse_app}, where application-level error and accelerator-specific \gls{ppa} metrics are used. ${BEHAV}_{\{\mathcal{O}_o \}}$ and ${PPA}_{\{\mathcal{O}_o \}}$ denote the application/task's behavioral error and the operator/accelerator's \gls{ppa} metrics when using the set of \glspl{axo} $\{\mathcal{O}_o \}$.
While Eqs.~\eqref{eq:dse_sel_syn} and~\eqref{eq:dse_app} show separate metrics for operator-level and application-specific optimization, \titleName~can be used to include metrics at different abstractions. For instance, the functional metrics at the task or application level can be used alongside operator-level \gls{ppa} metrics.
}

\begin{figure}[t] 
    \centering
       \scalebox{1}{\includegraphics[width=0.9 \columnwidth]{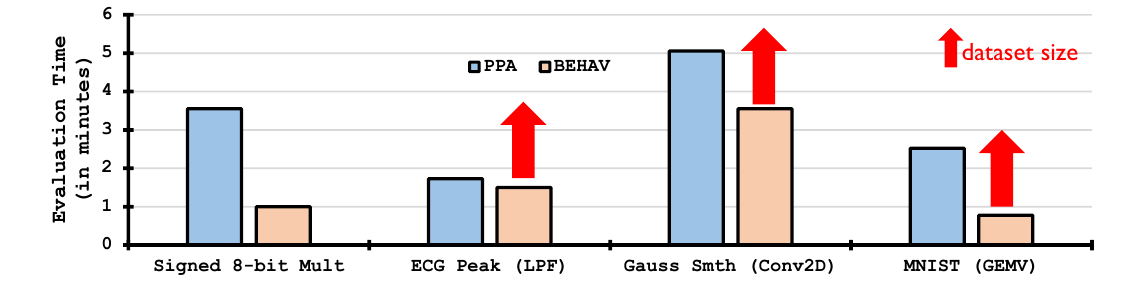}}
  \caption{Varying BEHAV and PPA estimation time for operator-level and application-specific exploration in~\cite{ullah2022appaxo}}
  \label{fig:est_ppa_behav} 
\end{figure}

\begin{equation}
\label{eq:dse_app}
\small{
\begin{split}
    \underset{\{\mathcal{O}_o \}}{\text{minimize}} &({BEHAV}_{\{\mathcal{O}_o \}},{PPA}_{\{\mathcal{O}_o \}}) \\
    s.t. ~ {BEHAV}_{\{\mathcal{O}_o \}} &\leq BEHAV_{MAX} ~~ and \\
    {PPA}_{\{\mathcal{O}_o \}} &\leq PPA_{MAX} 
\end{split}
}
\end{equation}

\siva{
The other aspect of \titleName-based \gls{dse} involves the \textit{Estimation} of the \gls{ppa} and \gls{behav} metrics.
The \gls{ppa} and \gls{behav} estimation methods are used in the operator-level and application-specific multi-objective optimization problems.
\titleName~allows researchers to integrate different estimation methods with varying accuracy and execution time trade-offs.
As can be seen in~\autoref{fig:arch_dse}, the estimation of the \gls{ppa} and \gls{behav} are abstracted as separate components with their interfaces to the rest of the \gls{dse} components.
Such an approach allows the researcher to use estimation methods of varying complexity and accuracy for each of the metrics and their combinations.
Further, as shown in~\autoref{fig:est_ppa_behav}, based on the data available in~\cite{ullah2022appaxo}, the estimation time for \gls{ppa} and \gls{behav} may vary considerably, depending on the granularity of analysis.
Also, the \gls{behav} estimation for application-specific \gls{dse} depends upon the dataset sizes used in the evaluation and may exceed the usually more time-consuming physical characterization for \gls{ppa} estimation.
Consequently, the integration of \gls{ml}-based estimation methods is necessary for any \gls{dse} framework for \gls{axc}.  
With increasing use of such surrogate fitness functions in iterative optimization methods, this enables integrating novel approaches to evaluating the effect of approximations efficiently by evaluating a large number of candidate solutions.
}

%% file: 40_tool_features.tex
\begin{figure*}[t]
    \centering
        \scalebox{1}{\includegraphics[width=0.95 \textwidth]{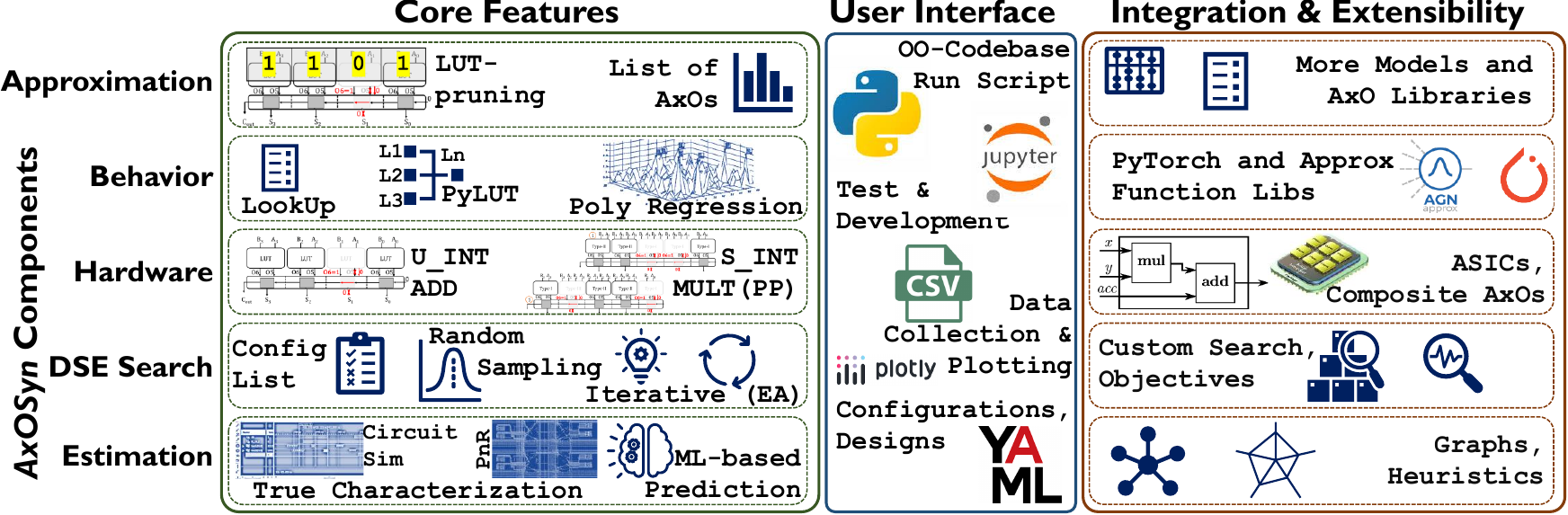}}
        \caption{Features of the proposed \titleName~framework}
    \label{fig:tool_features} 
\end{figure*}
\section{\titleName~Features}
\label{sec:tool_features}
\siva{
\titleName~is implemented as an object-oriented framework, providing a modular and extensible platform for exploring \glspl{axo}. 
The framework consists of various classes representing the main components, including \textit{functionality}, \textit{hardware}, \textit{approximation} models, \gls{dse} \textit{algorithms}, and methods for \gls{ppa}, and \gls{behav} \textit{estimation}. 
\autoref{fig:tool_features} provides a high-level overview of the proposed framework's features. 
The features of \titleName~can be categorized into \textit{core features} that are already implemented, \textit{interfaces} for using and extending the framework, and \textit{integration} and \textit{extensibility} features that allow the addition of new models and use cases. 
The core features include a robust set of implemented functionalities for synthesizing and evaluating \glspl{axo}. 
The framework also offers user-friendly interfaces for customization and extension, enabling researchers to add new approximation models, evaluation methods, or \gls{dse} algorithms. 
Additionally, \titleName~is designed to facilitate integration with other tools and use cases, making it a versatile and adaptable solution for exploring \glspl{axo} in various application domains. 
The features are described next in detail.
}

\input{41_core_features}
\input{42_interface}
\input{43_integ_extend}

%% file: 41_core_features.tex
\subsection{Core Features}

\subsubsection{Approximation Models and Operator Hardware/Behavior:}
\siva{
With the focus on enabling efficient \gls{dse} for \glspl{axo}, the proposed framework includes implementations of both AppAxO~\cite{ullah2022appaxo} or CoOAx~\cite{10.1145/3583781.3590222}-like models for generating novel \glspl{axo}~\cite{ullah2022appaxo,10.1145/3583781.3590222}. 
Both models use binary strings to selectively prune \glspl{lut} from accurate implementations, introducing deliberate approximations to balance \gls{behav} and \gls{ppa}. 
In addition, \titleName~provides abstractions for incorporating characterization data of \glspl{axo} from other models, such as EvoApprox~\cite{9218533}, which allows \gls{dse} using a selection-based method.
This abstraction allows users to experiment with a starting set of \glspl{axo} implementations instead of generating a new set of \glspl{axo} using an operator approximation model.
\titleName~is designed to model any arbitrary operator, but it currently includes specific modeling for signed integer multipliers and unsigned integer adders. 
The AppAxO model for these operators, with different operand bit-widths, is also integrated into \titleName. 
Three different behavioral modeling methods for \glspl{axo} are included: 
\begin{enumerate}
    \item The first method uses a lookup table of the results for all possible operand combinations.  The framework currently includes lookup-based models for EvoApprox~\cite{evoapprox16}.
    \item The second method encodes the functionality of each \gls{lut} used in the \gls{axo}'s implementation in software. The framework includes Python-encoded functionality for models such as AppAxO~\cite{evoapprox16} and CoOAx~\cite{10.1145/3583781.3590222}.
    \item The third approach employs \gls{pr}-based models to predict the approximate outputs of an \gls{axo} for given inputs. This approach was initially proposed by the authors of CLAppED~\cite{9586260} to predict the output behavior of any arbitrary \gls{axo} based on its error statistics.
\end{enumerate}
}

\siva{
These features related to approximation and behavior modeling enable efficient \gls{dse} of \glspl{axo} with varying quality versus search time trade-offs. 
For instance, the first and second methods provide the most accurate \gls{behav} estimation of using any arbitrary \gls{axo} in a task/application.
However, the first approach requires storing and looking up the result of every combination of inputs and may not scale well for wider bit-width operators.
Similarly, the second approach requires the user to provide a software model of the \gls{axo} hardware implementation, the execution of which can become very compute-intensive for complex hardware if the result is derived from event-driven simulations.
The third approach offers a much simpler approach, albeit with the possibility of errors in the estimation. 
While the originally proposed method in CLAppED~\cite{9586260} used a static method to the \gls{pr}-based estimation, \titleName~provides the parameterization of this approach.
\titleName~includes parameters for the degree of the polynomial and the number of random input combinations to be used for modeling the \gls{pr} expression.
This allows the user to model the behavior of each \gls{axo} without needing to simulate all possible combinations of the inputs.
}

\subsubsection{Design Space Exploration:}
\siva{ 
\titleName~includes implementations of various \gls{dse} search methods to cater to different needs. 
\begin{enumerate}
    \item The first method involves evaluating a list of predefined \gls{axo} configurations. 
    This allows the automated characterization of any arbitrary known set of designs. 
    Different use cases for this method include characterizing an initial set of designs for \gls{ml} models or a set of designs obtained from a \gls{dse} process.   
    \item The second method uses random sampling to explore different \gls{axo} configurations, relying on the random sampling mechanism implemented in the approximate operator's model.
    Integrating the random sampling mechanism into the operator model class allows the user to implement model-specific implementations.
    For instance, the random sampling for models such as AppAxO~\cite{ullah2022appaxo} and CoOAx~\cite{10.1145/3583781.3590222} requires generating a random binary string of an operator-specific length.
    On the other hand, for a selection-based method such as ApproxFPGAs~\cite{9218533}, random sampling involves picking from a library of existing designs, such as EvoApprox~\cite{evoapprox16}.
    \item The third method involves iterative search methods with multiple objectives, including both \gls{ppa} and \gls{ppa} metrics.
    To support multi-objective optimization, \titleName~includes an implementation of an evolutionary algorithms-based search method, \acrfull{ga}, allowing the exploration of trade-offs between different metrics effectively.
\end{enumerate}
}

\subsubsection{BEHAV and PPA Estimation:}
\siva{
The \gls{dse} feature of \titleName~integrates behavioral and hardware-related methods to facilitate the comprehensive exploration of \gls{axo} configurations.
For instance, the proposed framework includes classes for implementing various estimation methods for determining the resulting \gls{ppa} and \gls{behav} when using \glspl{axo}.
For hardware estimation, \titleName~provides physical characterization as well as machine learning-based regression models to predict \gls{ppa} metrics based on the approximation configuration. 
For functional characterization, in addition to using regression models, users can encode circuit simulation or other behavioral models of \glspl{axo} described above to estimate the \gls{behav} metrics. 
It must be noted that the \gls{behav} estimation being referred to here differs from the operator behavior estimation explained earlier.
By the ``operator behavior'' explained earlier, we refer to the modeling and estimation of the output of an \gls{axo} for any arbitrary input combination.
On the other hand, by ``\gls{behav} estimation'', we refer to the overall behavior of the operator/task/application when using any \gls{axo} and usually refer to statistical metrics such as \textit{average absolute error}, \textit{classification accuracy} etc.
}

%% file: 42_interface.tex
\subsection{User Interface}

\siva{
\titleName offers a Jupyter notebook-based interface designed to support both interactive experimentation and systematic design exploration. These notebooks provide a guided way for users to engage with the framework, enabling step-by-step interaction with each component—ranging from \gls{axo} modeling and synthesis to estimation and search. 
While primarily intended for development and testing, the notebooks can be adapted into Python scripts for running longer experiments or batch-mode evaluations. 
Configuration inputs are organized through YAML files, with Python dictionaries serving as the primary internal data structures for managing operator specifications, \gls{dse} parameters, and model configurations. 
Results and characterization data are logged using CSV files for downstream analysis and reproducibility.
}

\siva{
Each functional block within the framework is implemented as a Python class, derived from abstract base classes that define clear interfaces.
This modular structure allows users to selectively test, extend, or replace individual components, such as approximation models or search algorithms. 
The proposed framework's interface supports various workflows, including operator-level synthesis, accelerator characterization, and full \gls{dse} loops. 
For large-scale evaluations, multiprocessing support is integrated into key stages of the flow, particularly for \gls{behav} and \gls{ppa} estimation. 
Together, these features make the interface accessible for beginners yet flexible and powerful enough for advanced users exploring custom approximation workflows.
}

%% file: 43_integ_extend.tex
\subsection{Integration and Extensibility}

\siva{
\titleName~is built with an emphasis on flexibility, supporting integration with external tools and workflows while providing well-defined extension points across its modeling, search, and characterization layers. 
Integration capabilities include support for importing pre-existing approximate operator libraries such as EvoApprox~\cite{evoapprox16} and AppAxO~\cite{ullah2022appaxo}, which can be used for selection-based \gls{dse} or to initialize synthesis-driven exploration. 
The framework provides native compatibility with PyTorch-based inference pipelines, allowing application-level behavioral evaluation using real \gls{ml} workloads and datasets. 
Additionally, it accommodates composite or non-standard operators and allows targeting different \gls{fpga} families, enabling broader applicability in hardware design. 
The search process is customizable through interfaces for custom sampling routines, evolutionary search algorithms using DEAP~\cite{DEAP_JMLR2012}, and multi-objective optimization with user-defined objectives or constraints.
}

\siva{
In terms of extensibility, \titleName allows users to add new pruning techniques, generative \gls{axo} models, and even cross-layer approximations. 
It can be extended to support advanced workflows such as \gls{axat} and evaluation of \glspl{asic} alongside \glspl{fpga}. 
Extensions to \gls{mbo} and other forms of model-based search are supported by decoupling search logic from approximation models. 
Moreover, the modeling abstractions can be extended to support graph-based architectures such as \glspl{gnn}, allowing \titleName~to scale with emerging \gls{ai} algorithms and implement more complex operator models and estimation methods.
}

%% file: 50_expres.tex
\section{Experiments and Results}
\label{sec:expres}
\subsection{Design of Experiments}

\siva{
The experimental evaluation focuses on demonstrating the modularity, configurability, and extensibility of \titleName~through representative use cases. Instead of comparing performance against other tools, the goal is to validate the framework’s support for different approximation models, estimation approaches, and design space exploration strategies. Experiments are structured to cover operator-level modeling and exploration, hardware characterization, and application-level evaluation of approximate operators in end-to-end tasks.
All experiments were executed using the framework's Jupyter-based interface with YAML-driven configurations and Python dictionary-based data handling. Hardware characterization was performed using Vivado 2022.1~\cite{vivado}, targeting the Zynq-7000 SoC FPGA on the PYNQ-Z1 board. Design space exploration involved random sampling and evolutionary search using the DEAP~\cite{DEAP_JMLR2012} Python library. Multiprocessing was employed for scalable characterization, and results were visualized using Plotly-based interactive notebooks. Operator-level exploration was demonstrated on several types of adders and multipliers (e.g., Unsigned integer adders and signed integer multipliers of different bit-widths), and their design metrics (e.g., error probability, average absolute error, LUT and CC usage, critical path delay, power dissipation) were collected and visualized using Plotly-based interactive notebooks. 
}
\input{51_core_features}


\input{52_integ_exten}


\input{53_dse_axo_op}


%% file: 51_core_features.tex
\begin{figure*}[t]
    \centering
        \scalebox{1}{\includegraphics[width=1 \textwidth]{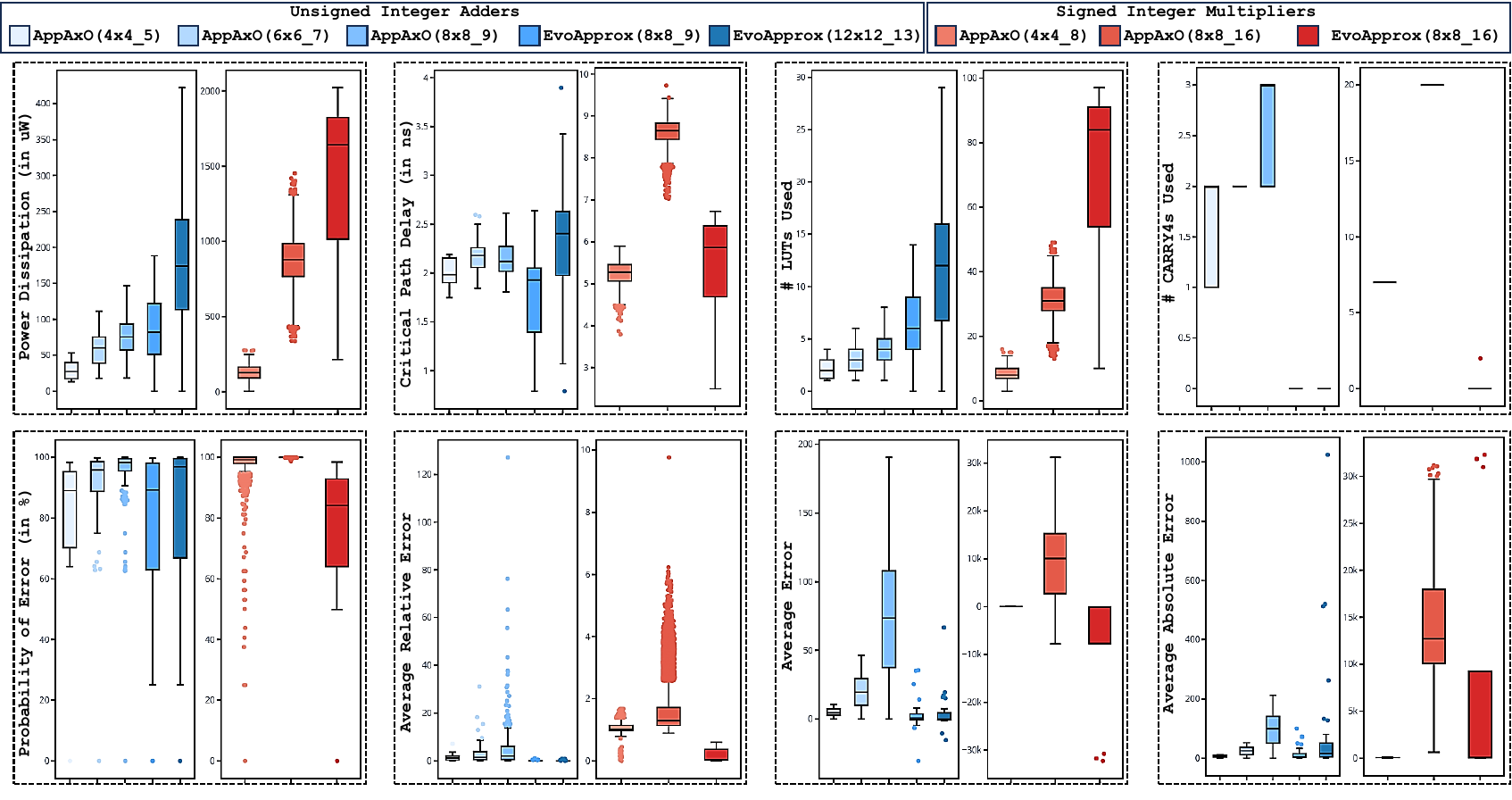}}
        \caption{Operator diversity in current version of \titleName. The exploration of unsigned integer adders and signed integer multipliers from AppAxO~\cite{ullah2022appaxo} and EvoApprox~\cite{evoapprox16} are integrated into the framework. Naming convention: 6x6\_7 indicates two input operands of 6 bits each and one output operand of 7 bits}
    \label{fig:exp_op_diversity} 
\end{figure*}
\subsection{Core Features}


\subsubsection{Approximation Diversity: Operators and Models:}
\siva{
To demonstrate \titleName’s support for multiple operator models and its ability to capture diverse design spaces, we compare representative subsets of approximate unsigned integer adders and signed integer multipliers generated using AppAxO (synthesis-based)~\cite{ullah2022appaxo} and EvoApprox (library-based)~\cite{evoapprox16} like methods. The box plots in~\autoref{fig:exp_op_diversity} illustrate the distribution of key \gls{behav} and \gls{ppa} metrics across different bit-widths and approximation strategies for both operator types.
}

\siva{
For unsigned adders, shown as blue-colored boxes on the left half of each sub-figure, AppAxO-based designs at increasing bit-widths (4-bit, 6-bit, 8-bit) exhibit a progressively wider spread in both behavioral error (e.g., Average Absolute Error and Probability of Error) and hardware cost metrics (e.g., \glspl{lut}, \gls{cpd} and power). This reflects the synthesis-driven nature of AppAxO, where a continuous design space is explored with varying levels of trade-off. In contrast, EvoApprox-based 8-bit and 12-bit adders show more discrete clustering of metric values, corresponding to their fixed library-based design entries. Despite this, EvoApprox still contributes a broad set of options at specific bit-widths, demonstrating its relevance in selection-based flows. It can be noted that the EvoApprox library usually includes designs that do not contain any logic functions and rely on the routing of input bits to the output to generate the operator's results. Hence, the lower minima values in the plots. Further, unlike EvoApprox, AppAxO relies on using the \gls{fpga}'s \glspl{cc} to reduce resource usage. As a result, EvoApprox designs show barely any CARRY4 usage and much higher \gls{lut} usage maxima.
}
\begin{figure*}[t]
    \centering
        \scalebox{1}{\includegraphics[width=1 \textwidth]{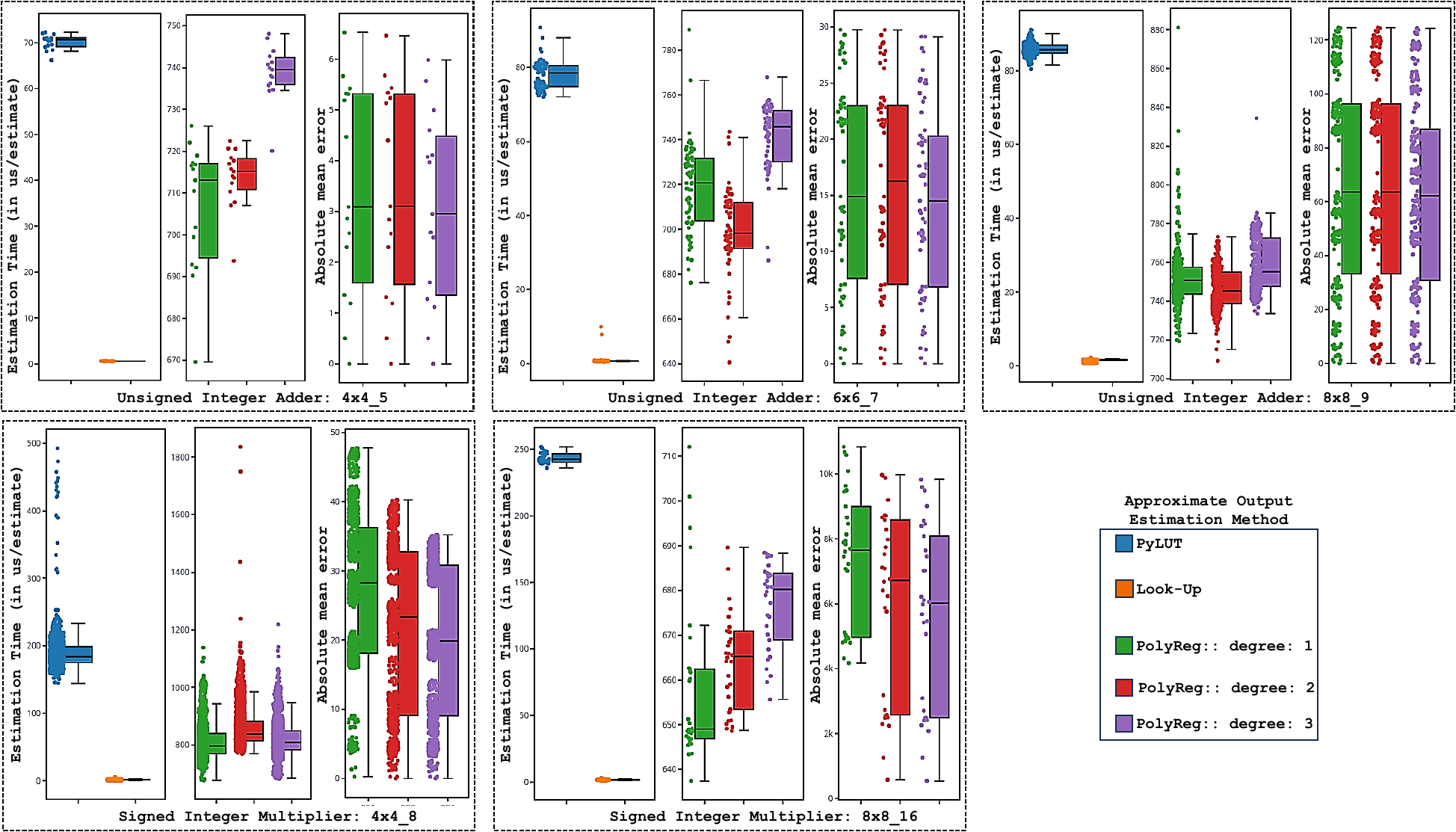}}
        \caption{Comparing various approximate output estimation methods in \titleName. The box plots show the distribution of the estimation time across different approximate operators of each operator type using an AppAxO~\cite{ullah2022appaxo} like approach. The distribution of errors using the polynomial regression-based estimation is also shown. Since \textit{PyLUT} and \textit{Look-Up} methods use the complete truth table, they do not result in any errors}
    \label{fig:exp_axout_est} 
\end{figure*}

\siva{
For signed multipliers, shown as red-colored boxes on the left half of each sub-figure, a similar trend is observed. AppAxO-generated 4-bit and 8-bit multipliers cover a wide range of accuracy and cost characteristics, while EvoApprox’s 8×8 multipliers cluster around specific high-error, low-cost designs. This highlights the difference in design philosophies--AppAxO enables generative exploration across a continuous approximation spectrum, whereas EvoApprox provides well-optimized discrete design points with known characteristics.
}
\siva{
Overall, this experiment demonstrates \titleName’s ability to ingest, represent, and explore both model types seamlessly. It also underscores the framework’s value in surfacing diverse trade-offs between approximation error and implementation cost—critical for downstream design space exploration and application-specific \gls{axo} selection.
}
\subsubsection{Operator Behavior Modeling:}
\siva{
To demonstrate the flexibility of \titleName's estimation pipeline, we compare the performance of several behavioral error modeling strategies across a diverse set of \gls{axo} instances. \autoref{fig:exp_axout_est} illustrates the distribution of estimation time and errors in the estimation of different methods: PyLUT, which uses \gls{lut}-level behavioral functions in Python to calculate the output, truth table look-up, and polynomial regression-based models of degrees 1, 2, and 3. These estimators are evaluated on multiple \gls{axo} types -- both unsigned adders and Baugh-Wooley-based~\cite{baugh1973two} signed multipliers, capturing the errors in estimation and computational costs.
}

\siva{
The results highlight key trade-offs in accuracy and generalization. PyLUT, which uses full functional simulation, provides consistent and accurate estimates but requires behavioral models of the approximate circuit implementation, making it suitable for offline or small-batch evaluations. Look-up table-based estimation, on the other hand, offers negligible latency but is limited to pre-characterized designs and lacks generalization to unseen configurations. Polynomial regression models offer a middle ground: first-degree models capture broad trends but exhibit higher variance in predictions, while higher-degree models improve accuracy at the cost of increased estimation time. The observed variability across operator types reinforces the value of the proposed framework's pluggable estimation interface, which allows users to choose or design estimators tailored to their performance and scalability needs.
}
\subsubsection{PPA and BEHAV Estimation:}
\begin{table}[t]
\centering
\caption{Comparing PPA and BEHAV estimation methods}
\label{table:ml_est}
\def\arraystretch{1.0}
\resizebox{1\textwidth}{!}{
\begin{tabular}{@{}ccccclcccllcc@{}}
\cmidrule(r){1-5} \cmidrule(l){7-13}
\multicolumn{5}{c}{\textbf{ML Modeling Accuracy}}                                                                                &                       & \multicolumn{7}{c}{\textbf{Characterization Time (in seconds)}}                                                                                                                                                                                                                                               \\ \cmidrule(r){1-5} \cmidrule(l){7-13} 
\multicolumn{1}{|c}{\textbf{Metric}}    & \multicolumn{2}{c}{\textbf{PDP}} & \multicolumn{2}{c|}{\textbf{AVG\_ABS\_ERR}}         & \multicolumn{1}{c|}{} & \multicolumn{3}{c}{\textbf{Signed Integer Multiplier 4x4\_8}}                                                         & \multicolumn{1}{c}{} & \multicolumn{3}{c|}{\textbf{Signed Integer Multiplier 8x8\_16}}                                                                                                \\ \cmidrule(lr){7-9} \cmidrule(l){11-13} 
\multicolumn{1}{|c}{\textbf{Operator}}  & \textbf{TRAIN}  & \textbf{TEST}  & \textbf{TRAIN} & \multicolumn{1}{c|}{\textbf{TEST}} & \multicolumn{1}{c|}{} & \textbf{\begin{tabular}[c]{@{}c@{}}PDP $\rightarrow$\\ AVG\_ABS\_ERR $\downarrow$\end{tabular}} & \textbf{PnR}                 & \textbf{PredML} & \multicolumn{1}{c}{} & \multicolumn{1}{c}{\textbf{\begin{tabular}[c]{@{}c@{}}PDP $\rightarrow$\\ AVG\_ABS\_ERR $\downarrow$\end{tabular}}} & \textbf{PnR}                 & \multicolumn{1}{c|}{\textbf{PredML}} \\ \cmidrule(r){1-5} \cmidrule(lr){7-9} \cmidrule(l){11-13} 
\multicolumn{1}{|l}{SINT MULT  4x4\_8}  & 23.76           & 66.43          & 0.179          & \multicolumn{1}{c|}{0.459}         & \multicolumn{1}{l|}{} & \multicolumn{1}{l}{\textbf{True Char}}                               & \multicolumn{1}{c|}{1189.51} & 2.395           &                      & \textbf{True Char}                                                                       & \multicolumn{1}{c|}{920.76}  & \multicolumn{1}{c|}{84.94}           \\ \cmidrule(lr){8-9} \cmidrule(l){12-13} 
\multicolumn{1}{|l}{SINT MULT  8x8\_16} & 467.15          & 855.37         & 254.69         & \multicolumn{1}{c|}{418.01}        & \multicolumn{1}{l|}{} & \multicolumn{1}{l}{\textbf{PredML}}                                  & \multicolumn{1}{c|}{1199.29} & 2.9876          &                      & \textbf{PredML}                                                                          & \multicolumn{1}{c|}{842.30} & \multicolumn{1}{c|}{4.116}          \\ \cmidrule(r){1-5} \cmidrule(l){7-13} 
\end{tabular}
}
\end{table}
\siva{
To demonstrate the utility of the proposed framework’s customizable \gls{ppa} and \gls{behav} estimation interface, we evaluated the performance of \gls{ml}-based predictors for both hardware and behavioral metrics against full synthesis-driven characterization. The comparison, summarized in \autoref{table:ml_est}, covers two signed integer multiplier configurations (4x4\_8 and 8x8\_16) and focuses on two key metrics: \gls{pdp} and average absolute error (AVG\_ABS\_ERR). For each case, we report model accuracy on training and test datasets, along with the time required to characterize a batch of operators using either the full synthesis and simulation flow or the \gls{ml}-based surrogate.
}

\siva{
The results show that \gls{ml}-based estimation can significantly reduce characterization time, especially for hardware metrics like \gls{pdp} that require full place-and-route analysis. While predictive accuracy is generally strong for smaller operators, larger and more complex designs exhibit higher test errors—particularly for behavioral metrics—due to the increased diversity in the approximation space. Importantly, these experiments serve as test cases using manually tuned models. In practice, \titleName~ allows users to integrate more sophisticated modeling pipelines, including AutoML~\cite{mljar} frameworks or domain-specific regressors, to improve prediction accuracy based on their application requirements. 
Further, abstracting the \gls{ppa} and \gls{behav} estimation into separate components allows the user to implement methods with differing cost and accuracy into the \gls{dse} methods. This is exhibited by the tables on the right half of~\autoref{table:ml_est}, where the processing time for varying combinations of methods is reported. The tables show the characterization time for 10 designs each. Also, the characterization for the 8x8\_16 operator uses two parallel threads.
This demonstrates the framework’s extensibility in estimation methodology and its ability to balance modeling fidelity with exploration speed.
}

%% file: 52_integ_exten.tex
\subsection{Integration \& Extensibility}

\begin{figure}[t] 
	\centering
    \subfloat[]{
        \centering
		\scalebox{1}{\includegraphics[width=0.5 \columnwidth]{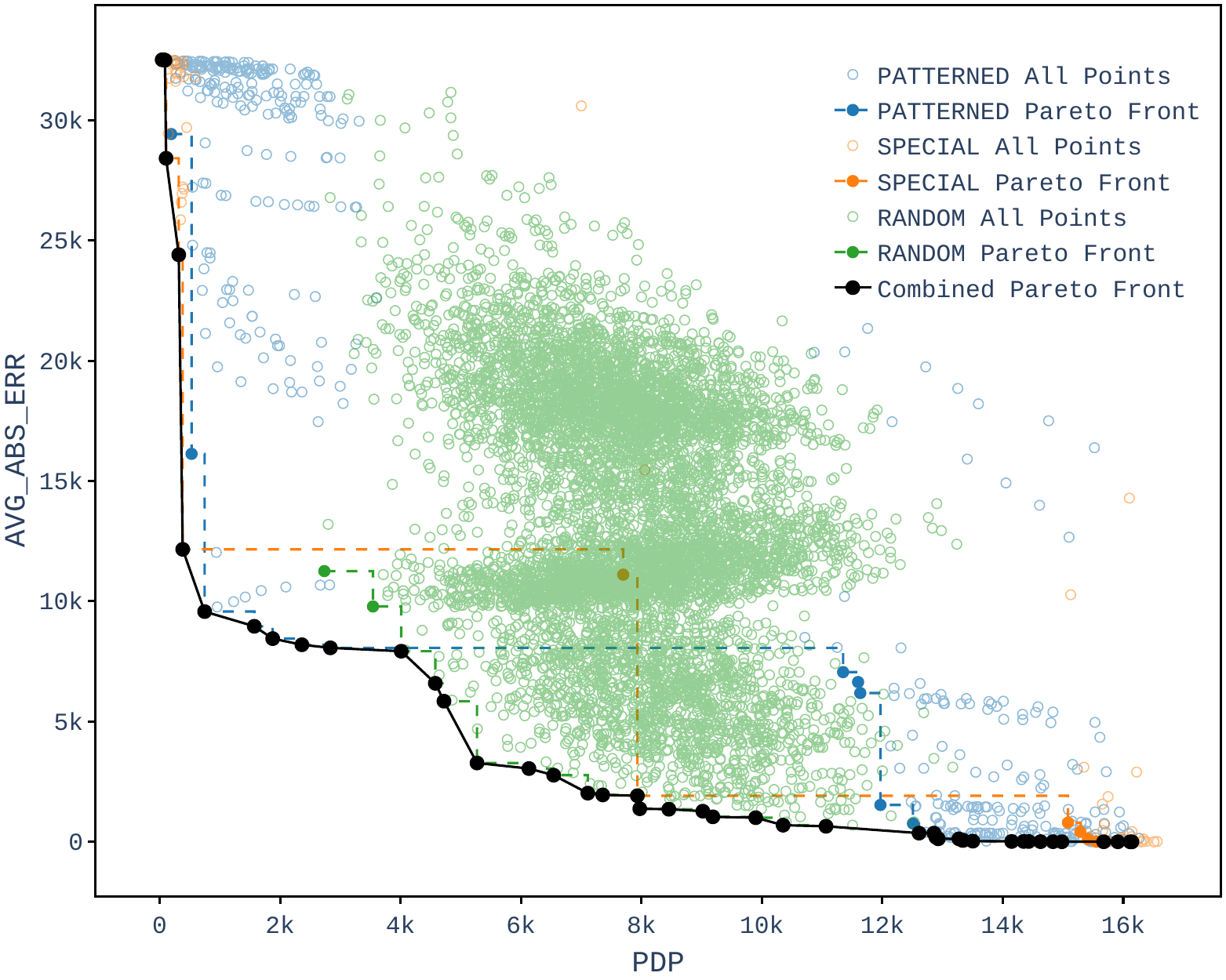}
	}}
	\subfloat[]{
		\centering
		\scalebox{1}{\includegraphics[width=0.5 \columnwidth]{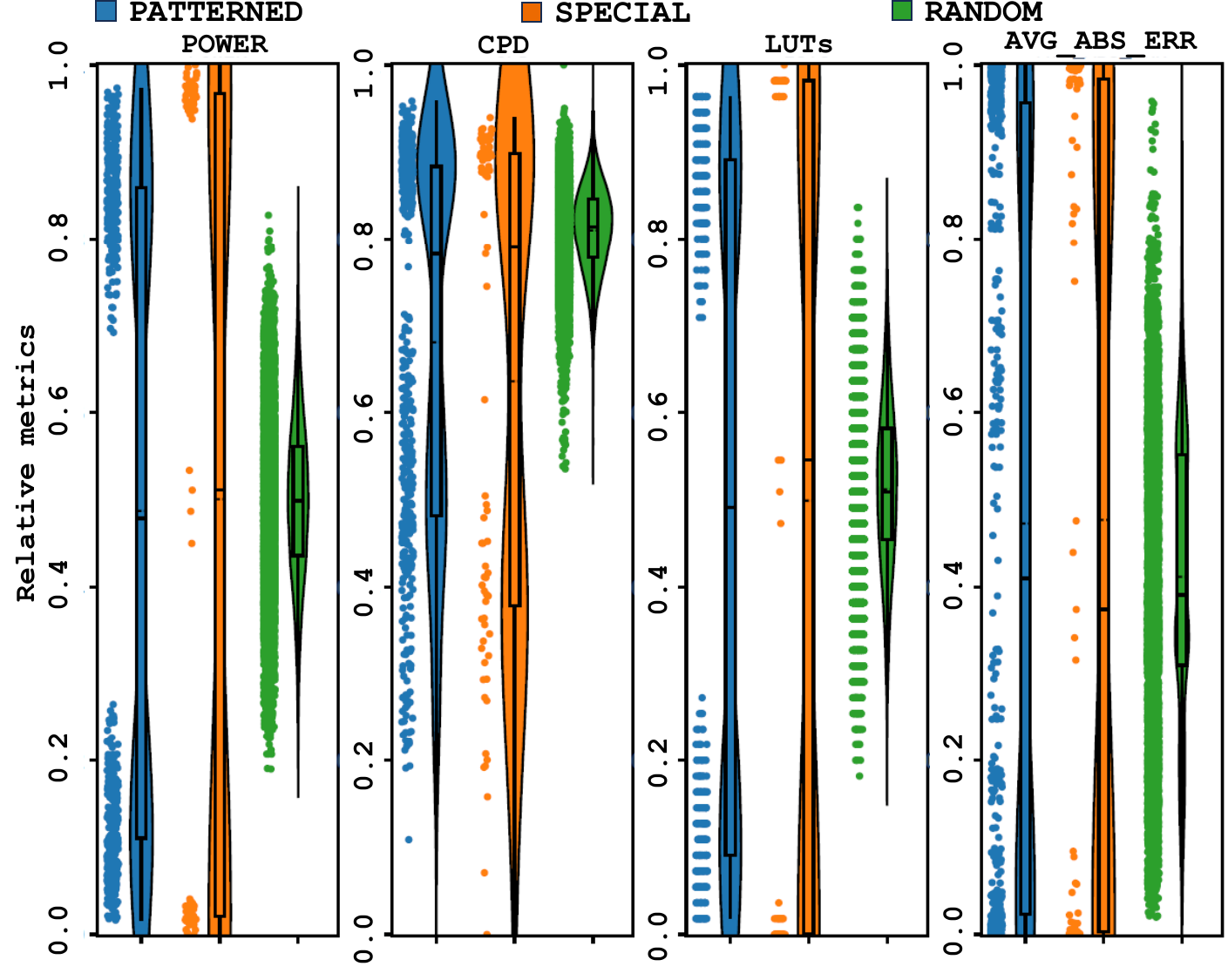}}
	}
	\caption{
    Comparison of the design points generated using different sampling techniques for an 8-bit signed integer multiplier. The AxO designs are generated based on the approach presented in~\cite{ullah2022appaxo} (a)~Scatter plot of all designs and the resulting Pareto front for each method. It also shows the combined Pareto front (b)~The distribution of PPA and BEHAV metrics for each of the three sampling methods
    }
	\label{fig:exp_sampling_var} 
\end{figure}
\subsubsection{Custom Sampling:} 
\siva{
To illustrate \titleName's extensibility in exploration strategy, we evaluated the effect of using user-defined sampling functions on the diversity and quality of approximate operator designs. \autoref{fig:exp_sampling_var} shows results from three different sampling modes implemented within the proposed framework for Baugh-Wooley signed integer multipliers (8x8\_16):
\begin{enumerate}
    \item \textit{RANDOM} sampling, where binary encodings of \gls{axo} designs generated with a synthesis approach are uniformly sampled
    \item \textit{PATTERNED} sampling, where structured windows of 0s and 1s sweep through the design space,
    \item \textit{SPECIAL}-pattern sampling, which includes handcrafted patterns such as alternating bits or single-bit activations.
\end{enumerate}
}

\siva{
The scatter plot in~\autoref{fig:exp_sampling_var}(a) highlights the differences in coverage and trade-off quality across the sampling modes. While RANDOM sampling broadly covers the space, the PATTERNED and SPECIAL-pattern strategies yield higher concentrations of low-power or low-error designs, with some configurations contributing uniquely to the combined Pareto front. The distribution plots shown in~\autoref{fig:exp_sampling_var}(b) further quantify this, showing how each sampling method favors different regions of the design space across key metrics like power, \gls{cpd}, LUT usage, and average absolute error.
}

\siva{
These results validate \titleName's ability to incorporate and evaluate non-standard sampling logic with minimal overhead, allowing users to inject domain knowledge or hypothesis-driven structures into the exploration process. Implementation-wise, the sampling method for generating new \gls{axo} designs is integrated into the approximation class definitions to allow model-specific customizations. For instance, the random sampling of graph-based models would vary significantly from the binary string-based sampling method shown above~\cite{ullah2022appaxo,10.1145/3583781.3590222}.
}



%% file: 53_dse_axo_op.tex
\subsection{Design Space Exploration}
\begin{figure}[t] 
	\centering
    \subfloat[]{
        \centering
		\scalebox{1}{\includegraphics[width=0.62 \columnwidth]{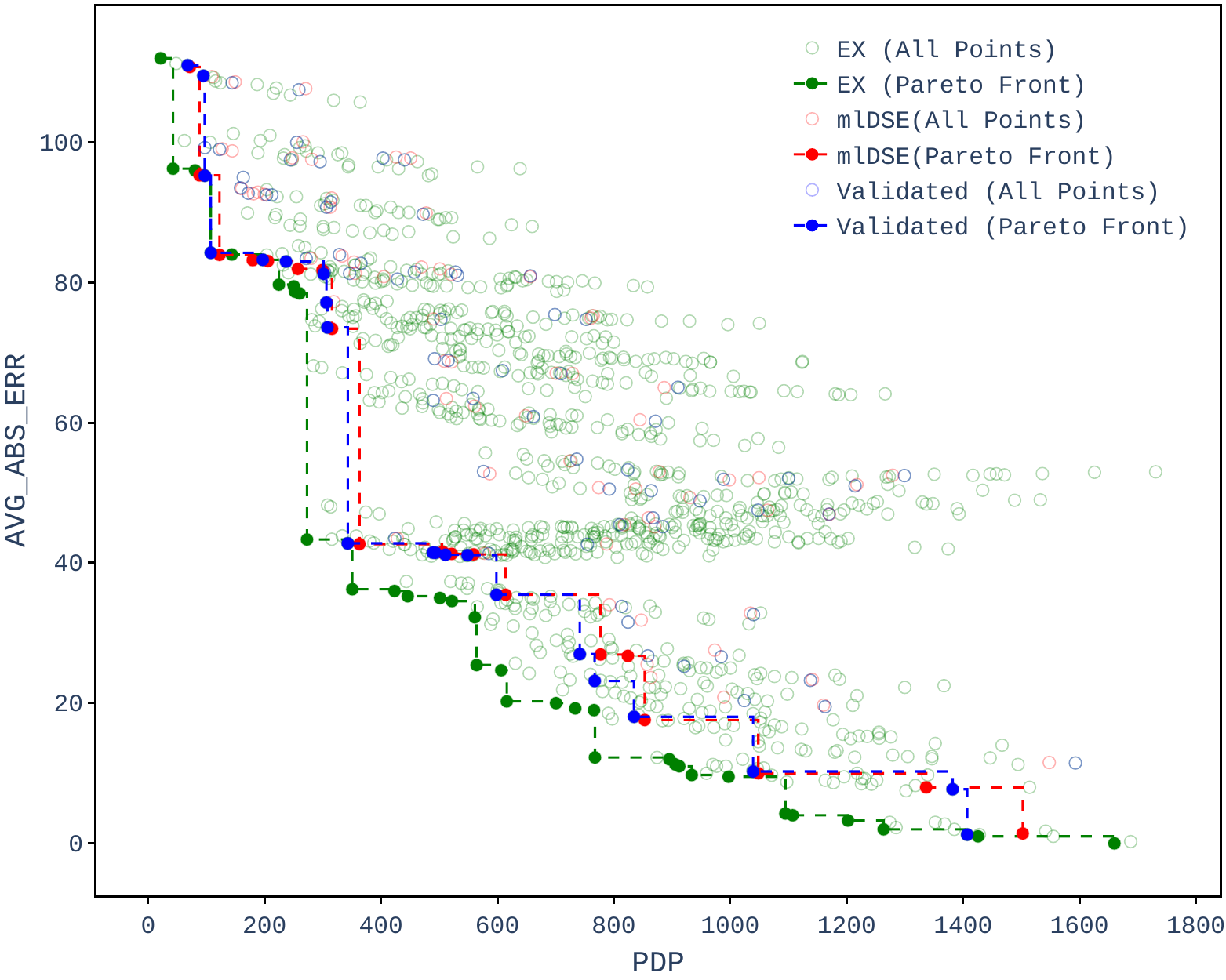}}
	}
	\subfloat[]{
		\centering
		\scalebox{1}{\includegraphics[width=0.38 \columnwidth]{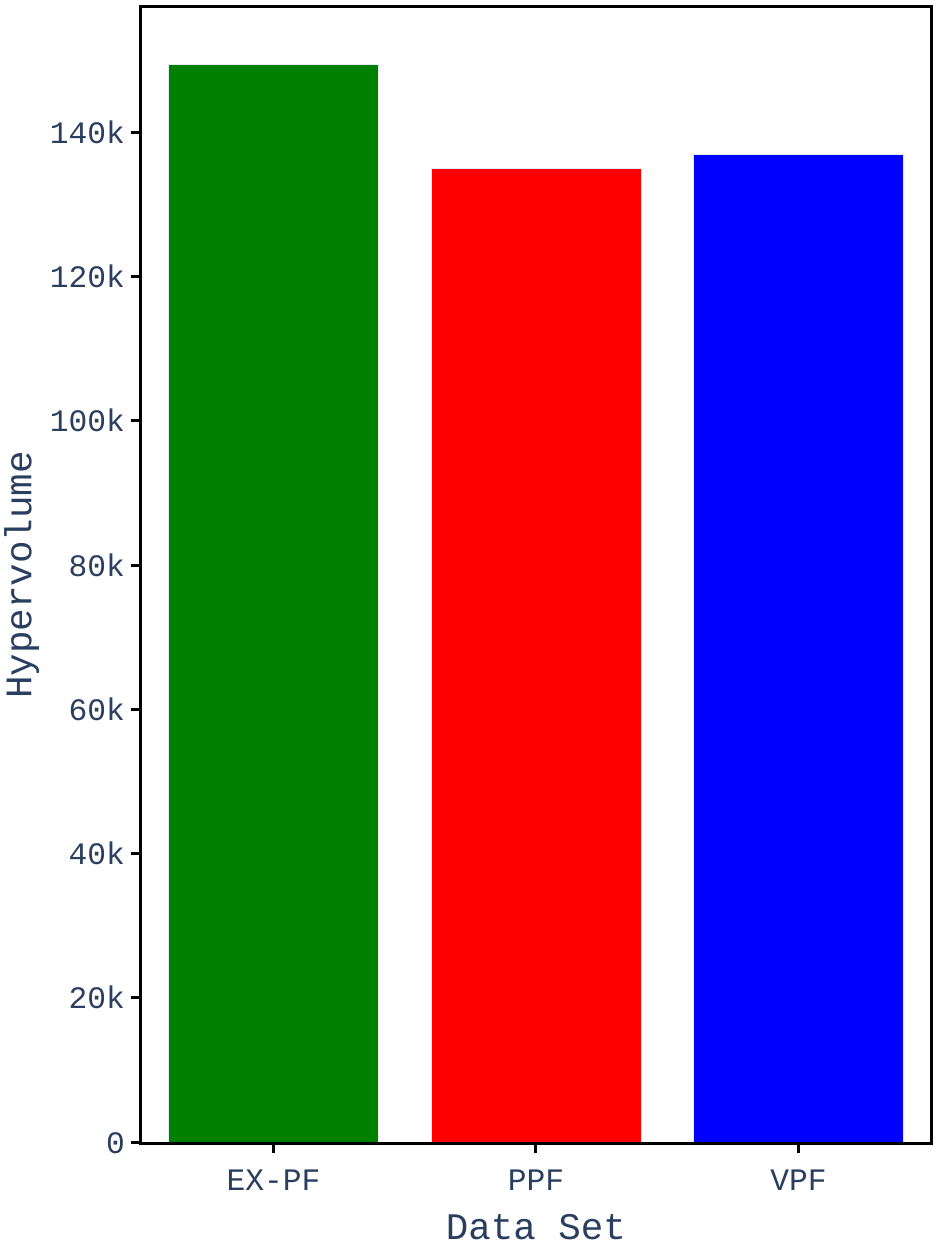}}
	}
	\caption{
    Comparison of the results from the GA-based DSE and the exhaustive design space for approximate implementations of signed 4-bit multipliers (a)~Comparing the Pareto fronts of different sets of designs (b)~Comparing the hypervolume of the Pareto front designs in each set 
    }
	\label{fig:op_dse} 
\end{figure}
\siva{
To highlight \titleName's support for \gls{dse}, we conducted an AppAxO-like synthesis-based generator experiment to explore the trade-offs in signed 4-bit integer Baugh-Wooley multiplier's approximation. This experiment demonstrates how \titleName~can integrate exhaustive enumeration, \gls{ml}-guided search, and full-characterization validation into a unified and extensible \gls{dse} workflow. The exploration targets a two-objective optimization: minimizing the \gls{pdp} as a proxy for hardware's \gls{ppa} and the average absolute error (AVG\_ABS\_ERR) as a measure of \gls{behav} metric due to approximation.
}

\siva{
\autoref{fig:op_dse}(a) presents a scatter plot of design candidates in the two-dimensional objective space (PDP on the x-axis, AVG\_ABS\_ERR on the y-axis), comparing three sets of designs:
\begin{itemize}
    \item The \textit{EX} set (green) represents the exhaustive sweep of all possible configurations generated using an AppAxO-like encoding space. Both the full set of points and the Pareto-optimal frontier are shown, illustrating the full range of possible trade-offs.
    \item The \textit{mlDSE} set (red) contains a subset of ~88 designs explored using a \gls{ga}-based search strategy within \titleName, where fitness evaluation is based on \gls{ml}-predicted estimates for both \gls{pdp} and AVG\_ABS\_ERR. This demonstrates the framework’s capability to incorporate learning-based models in guiding \gls{dse} without requiring full characterization at every step.
    \item The \textit{Validated} set (blue) includes the same 88 designs from the \gls{ga} search but re-evaluated using full place-and-route and simulation flows for \gls{pdp} and high-fidelity estimators for AVG\_ABS\_ERR. The re-characterization helps assess how well the \gls{ml}-guided search approximates the actual trade-off landscape.
\end{itemize}
The plot shows how different exploration strategies populate the objective space. While the exhaustive set provides the densest coverage, the \gls{ml}-guided search identifies a competitive Pareto front using a fraction of the evaluations, and the validated designs retain similar Pareto front characteristics, supporting the reliability of the surrogate models.
}

\siva{
\autoref{fig:op_dse}(b) complements the scatter plot with a hypervolume analysis, quantifying the quality of the Pareto fronts from each design set. The bars compare the hypervolume enclosed by the Pareto front from the exhaustive designs (EX-PF), the predicted Pareto front from the \gls{ml}-guided search (PPF), and the Validated Pareto front (VPF). Despite the significantly reduced evaluation cost, the \gls{ga}-driven exploration using predicted metrics yields a comparable hypervolume to the exhaustive set, and the validated designs confirm the robustness of the selected solutions.
Together, these results demonstrate \titleName’s ability to support multi-objective \gls{dse} workflows with flexible combinations of synthesis, predictive modeling, and post-hoc validation. The framework allows researchers to mix and match these components based on their exploration goals, available resources, and desired fidelity, making it a powerful tool for operator-level approximation studies.
}

%% file: 60_conc.tex
\section{Conclusion}
\label{sec:conc}
\siva{
In this work, we introduced \titleName, an open-source, modular framework that addresses critical limitations in the current landscape of \gls{dse} for approximate arithmetic operators. 
Existing frameworks are often either limited to selection-based approaches using static libraries or narrowly focused on fixed approximation models at specific abstraction levels. 
\titleName~bridges this gap by supporting both selection and synthesis-based design of \glspl{axo}, while offering a unified infrastructure for multi-objective optimization involving \gls{behav} and \gls{ppa} trade-offs. 
The framework tackles a pressing research challenge—how to systematically and efficiently explore the vast design space of approximate operators under application-specific constraints---by introducing support for customizable approximation models, layered modeling of hardware and application components, and flexible integration of physical or \gls{ml}-based estimators. 
Furthermore, \titleName's extensibility across operator types, estimation methods, and search strategies enables researchers to pose and solve new optimization problems without needing to re-implement the core framework.
}

\siva{
Moving forward, \titleName~ opens several avenues for future research at the intersection of approximate computing, design automation, and \gls{ml}-driven optimization.
The framework’s modular design enables extensions beyond arithmetic operators, such as approximate accelerators, memory structures, and communication primitives.
Future work could also integrate cross-layer approximation techniques, where arithmetic approximations are coordinated with quantization, pruning, and software-level error tolerance strategies to achieve more holistic optimizations. 
There is significant potential in expanding \titleName~to support \gls{axat} for neural networks, allowing the search for \glspl{axo} to co-evolve with model training.
Furthermore, integrating generative models, meta-learning, and Bayesian optimization can significantly enhance search efficiency in large design spaces. 
At the system level, extending the modeling hierarchy to support heterogeneous platforms, application workflows, and dynamic task mappings could make \titleName~a central tool in approximate system co-design. 
By releasing \titleName~as an open and extensible research platform, we aim to enable further innovation in energy-efficient \gls{ai} system design by enabling the community to explore novel, application- and platform-aware approximation strategies.
}